\documentclass[twocolumn,prb,floatfix,showpacs]{revtex4}
\usepackage{graphicx,color}
\usepackage{subfigure}
\usepackage{verbatim}
\usepackage{ifpdf}
\usepackage{ifthen} 

\ifpdf
\usepackage[pdftex]{pdflscape}
\usepackage[pdftex,colorlinks=true]{hyperref}

\else
\usepackage{lscape}
\usepackage[dvips,colorlinks=false]{hyperref}

\fi

\hypersetup{
pdftitle={Grain boundary energies and cohesive strength as a function of geometry},
pdfauthor={Valerie R. Coffman, James P. Sethna},
pdfkeywords={grain boundaries, molecular dynamics, fracture mechanics, finite
  elements, solid state physics,
  computational physics},
bookmarksnumbered,
pdfstartview={FitH},
pdfpagemode=UseOutlines,
breaklinks=true
}

\newcommand{\mycaption}[2][]{
  \ifthenelse{\equal{}{#1}} {\caption{#2}} {\caption[#1]{ {\bf#1} #2}}}

\newcommand{\inThesis}[1]{}
\newcommand{\inPaper}[1]{#1}
\newcommand{\figWidth}{8cm}
\newcommand{\coloronline}{}

\begin{document}

\title{Grain boundary energies and cohesive strength as a function of geometry}

\author{Valerie R. Coffman, James P. Sethna}
\affiliation{Laboratory of Atomic and Solid State Physics (LASSP), Clark Hall,
Cornell University, Ithaca, NY 14853-2501, USA}

\date{\today}

\begin{abstract}
Cohesive laws are stress-strain curves used in finite element calculations to
describe the debonding of interfaces such as grain boundaries.  It would be
convenient to describe grain boundary cohesive laws as a function of the
parameters needed to describe the grain boundary geometry; two parameters in 2D
and 5 parameters in 3D.  However, we find that the cohesive law is not a smooth
function of these parameters.  In fact, it is discontinuous at geometries for
which the two grains have repeat distances that are rational with respect to one
another.  Using atomistic simulations, we extract grain boundary energies and
cohesive laws of grain boundary fracture in 2D with a Lennard-Jones potential
for all possible geometries which can be simulated within periodic boundary
conditions with a maximum box size.  We introduce a model where grain boundaries
are represented as high symmetry boundaries decorated by extra dislocations.
Using it, we develop a functional form for the symmetric grain boundary
energies, which have cusps at all high symmetry angles. We also find the
asymptotic form of the fracture toughness near the discontinuities at high
symmetry grain boundaries using our dislocation decoration model.

\end{abstract}


\maketitle

\section{Introduction}

\begin{figure}[thb]
\begin{center}
\includegraphics[width=\figWidth]{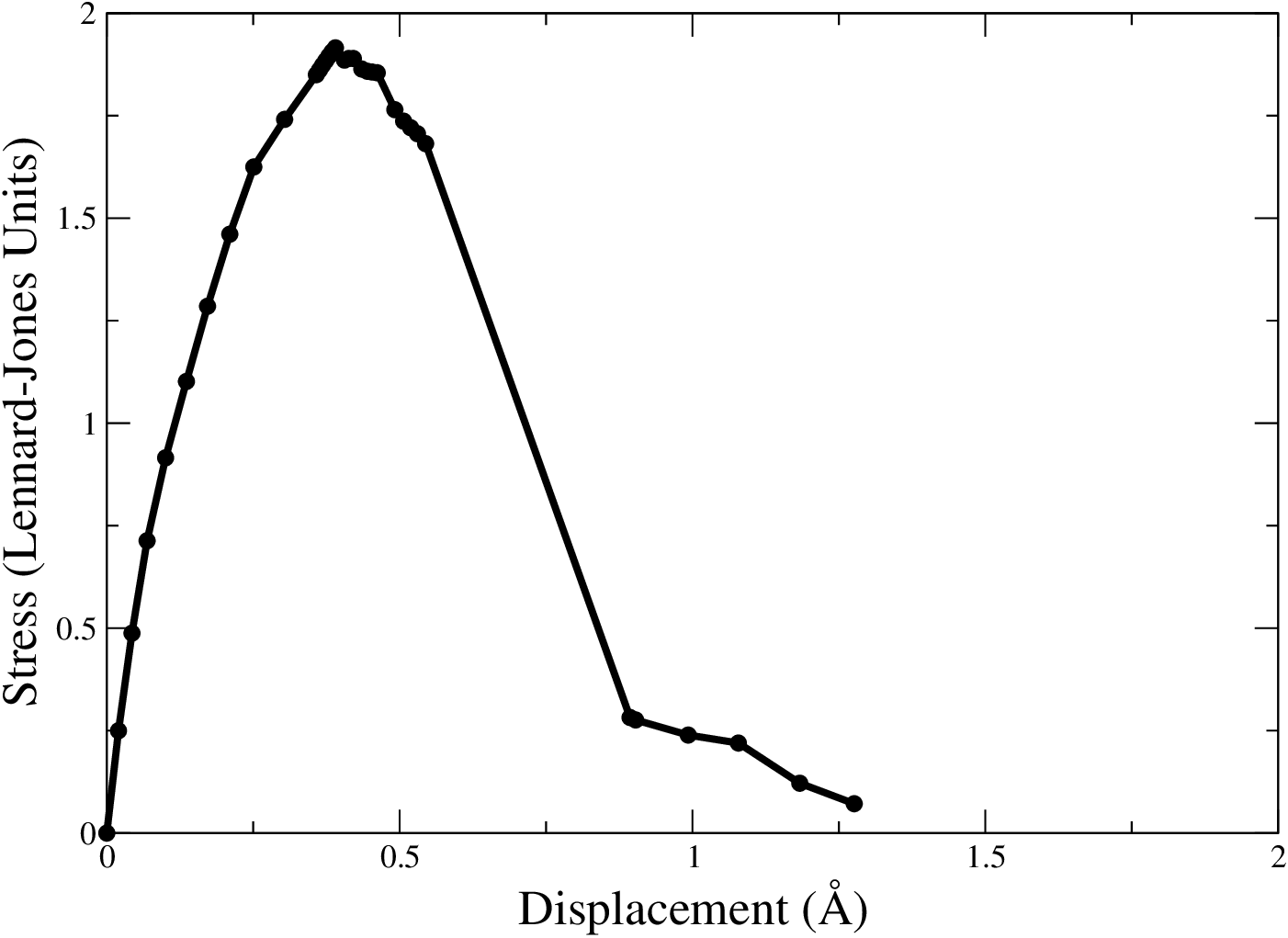}
\end{center} 
\mycaption[An Example of a Cohesive Law.]{ The stress vs.\ strain curve that
describes the debonding of a 2D grain boundary with tilt angles 33.418$^{\circ}$
and 26.58$^{\circ}$.  Our measurements include the elastic response of the
perfect crystal on either side of the grain boundary.  In the results shown
above, we have subtracted off the elastic response of the bulk in order to
isolate the elastic response of the interface. 
\inPaper{ This is described in
detail in~\cite{cube-in-cube}} 
\inThesis{ This is described in detail in
Chapter~\ref{chapter:cube_in_cube}} 
where cohesive laws of grain boundaries are
used in finite element simulations of polycrystals.
%For input
%into finite element simulations, which need the cohesive law of only the
%interface, we must subtract off the elastic response of the bulk as shown
%here. \inThesis{ This is described in detail in
%Chapter~\ref{chapter:cube_in_cube}.} \inPaper{ This is described in detail
%in~\cite{cube-in-cube}.}
}
\label{fig:CohesiveLaw}
\end{figure}

%The initiation of fracture in a material is dominated by the microstructure and
%defects of the crystal lattice [cite sources].  
In materials such as silicon and various aluminum alloys, cracks initiate and
propagate along the interfaces between polycrystals known as grain boundaries.
When the cracks initiate at the site of these microscopic defects, the
macroscopic fracture strength of the material is dependent on the
microscopic structure of the grain boundaries.  The debonding of
an interface such as a grain boundary is described by a cohesive law, giving the
displacement across the interface as a function of stress (figure~\ref{fig:CohesiveLaw}).
%such as
%the one shown in figure~\ref{fig:CohesiveLaw}.
%Such laws are used in finite element modeling of
%polycrystals for the initiation of fracture.   

Cohesive laws are used by finite element cohesive zone models, which simulate
fracture initiation at interfaces~\cite{needleman-interfaceCZM}.  It has been
shown that the shape and scale of the cohesive law has a large effect on the
outcome of the finite element simulation~\cite{needleman-interfaceCZM,
falk-critical}. However, the CZM studies of grain boundary fracture have used
cohesive laws that are guessed, chosen for numerical convergence, and do not
take into consideration the effect of varying grain boundary geometries within
the material -- the same cohesive law is often used throughout the material
despite the fact that in a real material, grain boundaries of varying geometries
must occur~\cite{iesulauro1, iesulauro2, warner-sliding-plasticity}.

It would be useful to find a formula for the cohesive laws of the grain
boundaries of a given material as a function of geometry, for input into finite
element simulations.  The geometry of a 3D grain boundary depends on 5
parameters that describe the orientations of the two grains.  In addition, there
are three different modes of fracture (normal to the crack plane, shear in a
direction parallel to the crack line, or shear in a direction perpendicular to
the crack line) to explore~\footnote{For cracks propagating through a grain
boundary, the direction of the crack front in relation to the structure of the
grain boundary may affect the fracture toughness, but this dependence is not
included in cohesive laws.} as well as dependencies on temperature, impurities
at the interface, and emission of dislocations to consider.

Thus far, no systematic study of grain boundary cohesive laws as a function of
geometry has been done with molecular dynamics or experiment.  There is
difficulty in measuring cohesive laws experimentally because it is
difficult to isolate and measure the displacements on either side of
the grain boundary.

%Data for this plot is located on ccmr in
%ASP/MDWebServices/GrainBreaker/SymmetricGrainBoundaries/SymmetricUnder20Unique/geo106

% LITERATURE REVIEW OF PREVIOUS WORK COMBINED WITH OUTLINE OF PAPER/CHAPTER
% small number of 3D gb's
Previous atomistic studies of the mechanical response of grain
boundaries have concentrated on a small number of symmetric grain boundaries in
3D~\cite{ chen-GB,
sansoz-molinari-shear-tension,sansoz-molinari-structure,shenderova-diamond,
spearot-atomistic, warner-sliding-plasticity}.
%%%%%% 
Exploring the complete picture of 3D grain boundary cohesive laws involves
exploring a 5-dimensional space with three modes of fracture.  Because of this
difficulty, we have taken a step back.  We seek to systematically explore the
cohesive laws for mode I fracture for all possible grain boundary geometries
in 2D that can be simulated in periodic boundary conditions for a particular
size and strain. We will initially be focusing on symmetric grain boundaries and
then expanding the picture to look at asymmetric 2D grain boundaries.  One
use of such data would be for finite element simulations of polycrystals~\cite{cube-in-cube}, as
described above.  For this purpose, it would be nice to find a functional form
which describes the fracture strength of the grain boundary as a function of
geometry.

We find, however, that the fracture strength as a function of tilt angles is
discontinuous everywhere, with particularly large jumps at special, high
symmetry grain boundaries composed of a simple arrangement of structural units
with a low repeat distance. These special boundaries are also associated with
cusps in the grain boundary energy.  We will describe dependence of the energy
and the fracture strength near these special boundaries by treating them as
perfect, albeit complex, crystals with added dislocations to break the symmetry.

%csl
Most studies use the coincidence site model~\cite{bollmann} to construct and
classify special grain boundaries~\cite{chen-GB,
sansoz-molinari-shear-tension,sansoz-molinari-structure,shenderova-diamond,
warner-sliding-plasticity} while a few men\-tion the sig\-nif\-i\-cance of struc\-tural
units~\cite{bishop-chalmers,chen-GB,bubble-raft,sansoz-molinari-structure,wolf-merkle}.
% structure
The only 2D study we are aware of, the ``bubble-raft'' model, observes the
structural units of several of the special, high angle grain boundaries for the
triangular lattice and how patterns of structural units combine to create
vicinal geometries~\cite{bubble-raft}. Other 3D studies discuss how grain
boundaries with geometries close to those of special grain boundaries can be decomposed into the structural
unit of the special grain boundary with added flaws~\cite{bishop-chalmers,
wolf-merkle}. We develop a systematic way of finding high symmetry geometries
and show that the combinations of patterns of structural units at vicinal grain
boundaries are key to understanding the dependence of energy and fracture
strength on geometry.

% shift to minimize
Sansoz and Molinari find the grain boundary energy by allowing the
grains to relax together from an initial separation of a few angstroms
\cite{sansoz-molinari-structure} while others perform
a conjugate gradients search~\cite{chen-GB}.  We use a systematic method for
explicitly imposing a relative shift between the grain and using  atomistic
relaxation for finding the global energy minimum similar to that used in
\cite{sansoz-molinari-shear-tension, wolf-merkle}.
%energy cusps
It is well established that there exist cusps in the grain boundary energy
for special high angle grain boundaries~\cite{chen-GB,
old-grain-boundary-sims, pumphrey, sansoz-molinari-structure,
shenderova-diamond, wolf-structure, wolf-merkle, wolf-jaszczak}.
%shear tension
Recent studies of grain boundary constitutive properties focus on the response
to shear~\cite{sansoz-molinari-structure, warner-sliding-plasticity} or compare
shear and tension~\cite{sansoz-molinari-shear-tension, spearot-atomistic}.
Warner et al.~claim that the tensile response does not depend on the geometry
of the grain boundary~\cite{warner-sliding-plasticity}.
%jumps in fracture strength
Others have seen jumps in the tensile fracture strength at high angle grain
boundaries geometries~\cite{chen-GB, shenderova-diamond}.  

We will explain the cusps in energy and jumps in fracture strength by drawing an
analogy between perfect crystals and high symmetry, high angle grain boundaries.
The dislocation model of low angle grain boundaries gives a $\theta \log \theta$
form for the grain boundary energy.  Because vicinal grain boundaries can be
thought of as high symmetry grain boundaries with added flaws, the energy cusps
at special grain boundaries have the same $\theta \log \theta$ form. A similar
argument will apply to the peak fracture stress.  Just as adding dislocations
to the perfect crystal adds a nucleation point for fracture, and therefore a
discontinuity (jump down) in fracture strength, adding a flaw to a high
symmetry, high angle grain boundary abruptly changes the local fracture strength
by adding a potential nucleation site for fracture.

\section{Procedure for Calculating Grain Boundary Energy and Cohesive Strength}
\label{sec:measuring_cohesive_laws}

%basic info

We calculate energies and cohesive strengths for grain boundaries using
 the \textit{DigitalMaterial}\cite{DigitalMaterial} package to perform atomistic simulations.
 The potential we are using for 2D simulations is the
Lennard Jones potential with a smooth, fourth order cutoff between 2.41 and 2.7
\mbox{\AA}. The ground state is a triangular lattice with a lattice constant of
1.11 \mbox{\AA}.

% Data and original in
% /home/val/work/DataStore/SymmetricGrainBoundaries-6/Under20NonPerfectNonSymm/geo117/
\begin{figure}[thb]
\begin{center}
\includegraphics[width=\figWidth]{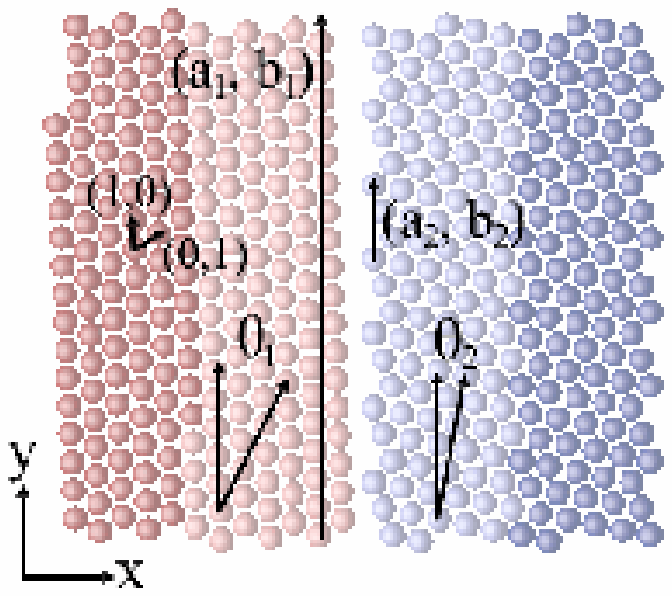}
\end{center}
\mycaption[An Example of a Grain Boundary.]{\coloronline The tilt angles are given by
$\theta_1$ and $\theta_2$ and the repeat distances are $D_1=|(a_1,b_1)|$ and
$D_2=|(a_2,b_2)|$.  We have assumed the convention where 0$^{\circ}$ indicates
an orientation with Miller indices $(2,1)$ for the vertical face of the grain
boundary.}
\label{fig:Geometry}
\end{figure}

We set up the grain boundary by initializing two rectangular grains with the
given rotations that define the grain boundary geometry we wish to measure.  An
example of the configuration of atoms that make up a grain boundary simulation
is shown in figure~\ref{fig:Geometry}.  We use periodic boundary conditions in
the $y$-direction in order to avoid edge effects which could potentially cause
cracks to nucleate at the intersection with the surface~\footnote{ Another
alternative  is to use a layer of constrained atoms.
Typically, in a finite element simulation of a subsection of a material, the
nodes on the surfaces of the model are constrained to not move in a direction
perpendicular to the surface.  This is known as ``rollered'' boundary conditions.
We can imitate these boundary conditions in an atomistic simulation by
constraining a layer of atoms at the surface or edge of the simulation to not
move perpendicular to that surface or edge.  This will partially ameiliorate the
edge effects, suppress the
Poisson effect,  and has the advantage of allowing us to simulate grain boundaries
of any geometry~\cite{cube-in-cube}. We have compared simulations of
different sizes with rollered boundary conditions in the $y$-direction to a
simulation of the same geometry, but with periodic boundary conditions in the
$y$-direction. We see that if we use rollered boundary conditions, we
need a simulation that is 48 times longer in the $y$-direction in order to
converge to within 2\% of the results found by a simulation with periodic
boundary conditions.  Hence, for computational efficiency, we choose to use periodic
boundary conditions in the $y$-direction.}.  This has the
disadvantage of only allowing geometries that have finite repeat distances. A
method for finding these geometries is discussed in detail in
section~\ref{sec:AllPossibleGeometries}. If necessary, both
grains are strained slightly by equal and opposite amounts in order to have both grains
fit into a periodic box.  We use a constrained layer of atoms to
impose fixed boundary conditions in the $x$-direction.  These are represented by
the darker atoms in figure~\ref{fig:Geometry}.  The constrained layer of atoms
has a width equal to twice the cutoff distance of the potential to eliminate
surface effects from the free atoms.

%\subsection{Finding the Most Natural Grain Boundary Configuration}
%\label{sec:Minimize}
Besides the tilt angles, there are other factors to consider in constructing the
grain boundary geometries. For commensurate grain boundaries, there will be an
ideal relative displacement in the direction parallel to the grain boundary (the
$y$-direction in figure~\ref{fig:Geometry}).  This ideal displacement will
correspond to the lowest energy and thus is the most natural configuration for a
given pair of tilt angles.~\footnote{ The Frenkel-Kontorova model
\cite{frenkel-kontorova} describes a one dimensional chain of atoms, connected
by springs, subject to an external, sinusoidal potential.  If the
relaxed length of the spring and the width of the potential wells are
commensurate a minimum energy state exists.  For the incommensurate case, there
will be a pinned phase with many local minima and an unpinned phase, depending
on the depth of the wells in relation to the stiffness of the springs.}

\begin{figure}[thb]
\begin{center}
\includegraphics[width=\figWidth]{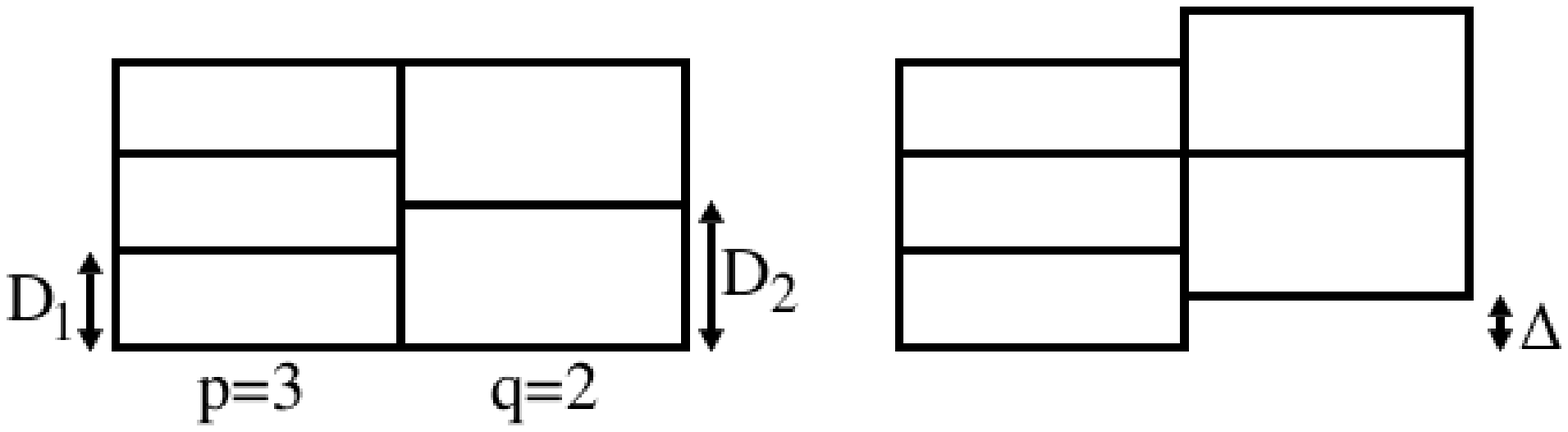}
\end{center}
\mycaption[The Necessary Search Range for $y$-displacements.]{ The minimum range we
must search over to find the most natural configuration is equal to the minimum
shift that gives an equivalent configuration of atoms along the surface.
Assuming that the edges of the repeat cells line up at the bottom, the minimum
shift to produce an equivalent configuration is done by lining up the top edge
of the first repeat cell in the grain with a larger repeat (the grain on the
right in the diagram) with the nearest edge of a repeat cell in the grain with a
smaller repeat distance (the grain on the left).  If $D_2$ is equal to an
integer number of repeats of $D_1$, the minimum search range is equal to $D_1$.
$\lfloor D_2/D_1 \rfloor$ gives the number of repeats of the left grain that fit
within a single repeat of the right grain and $\lceil D_2/D_1 \rceil$ gives the
number of repeats of the grain on the left that contain one repeat of the grain
on the right.  If $D_2/D_1$ is not an integer, the minimum search range is then
$\min \left(\left| D_2 - \lfloor D_2/D_1 \rfloor D_1 \right|, \left| D_2 -
\lceil D_2/D_1 \rceil D_1 \right| \right)$ }
\label{fig:PandQ}
\end{figure}

\begin{figure}
\begin{center}
\includegraphics[width=\figWidth]{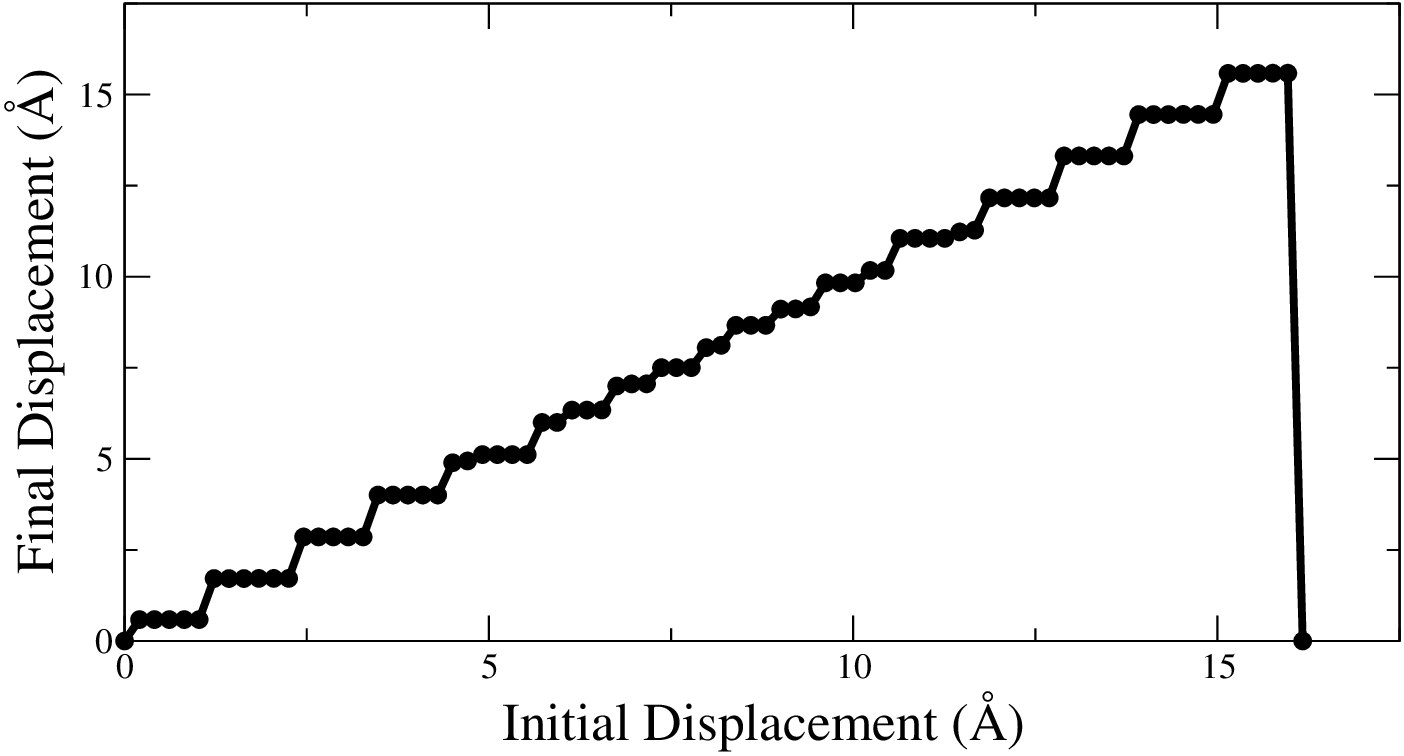}
\vskip .5 cm
\includegraphics[width=\figWidth]{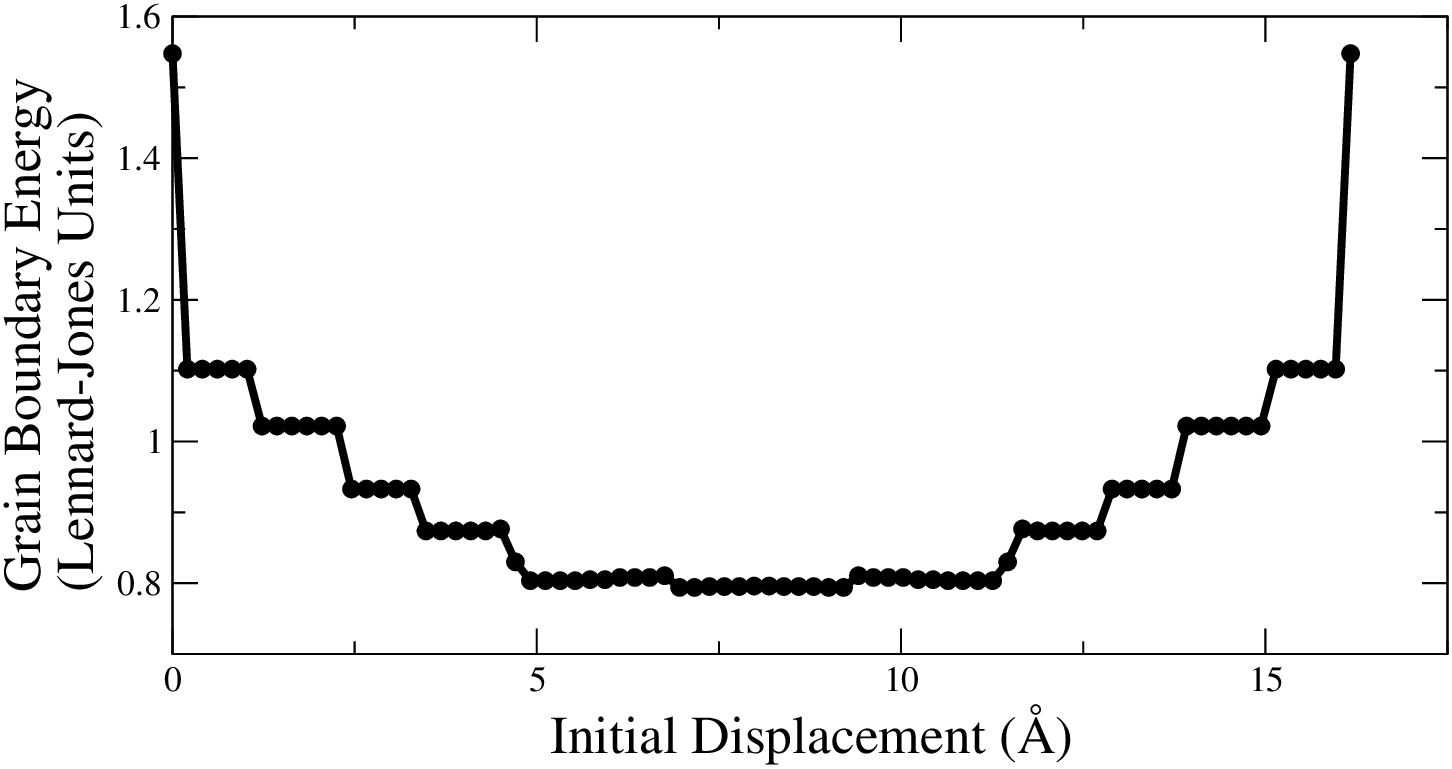}
\end{center}
\mycaption[Finding the Energy Minimizing Grain Boundary Configuration.]{ Each
plateau in the top plot and each flat region in the bottom plot correspond to
the basin of attraction of a local minimum.  The most natural grain boundary
configuration corresponds to the global energy minimum.}
\label{fig:MinSearch}
\end{figure}

%In setting up a grain boundary geometry for measurement of the cohesive law, we
%must find the displacement along the boundary which gives the lowest energy.  
We find the displacement along the boundary which gives the lowest energy by
initializing the two grains with a small displacement in the $x$-direction and
varying displacements in the $y$-direction, relaxing the atoms (with the
boundary layers contrained to be rigid and non-rotating), and measuring the
grain boundary energy per length.  The range of displacements in the
$y$-direction that we must search over is given by
\begin{equation}
\Delta = \left\{ \begin{array}{ll}
D_1 & D_2 \bmod{D_1} = 0 \\
\min \left( \left| D_2 - \lfloor D_2/D_1 \rfloor D_1 \right|, 
\inPaper{\right. & \\ \left. }
\left| D_2 -
\lceil D_2/D_1 \rceil D_1 \right| \right)  & D_2 \bmod{D_1} \ne 0
\end{array} \right.
\label{eqn:Delta}
\end{equation}
where we assume $D_1 < D_2$.  This is illustrated in figure~\ref{fig:PandQ}.
The grain boundary energy per length
is defined as
\begin{equation}
E_{GB} = \frac{E_{total} - N_{atoms} * E_{bulk}}{L}
\label{eqn:GrainBoundaryEnergy}
\end{equation}
where $E_{total}$ is the total potential energy for the configuration of atoms
(excluding the constrained atoms), $N_{atoms}$ is the number of unconstrained
atoms, $E_{bulk}$ is the energy of a single atom in the bulk, and $L$ is the
length of the grain boundary. An example of the results of such a search is
shown in figure~\ref{fig:MinSearch}.  The regions with the same final displacement
and energy correspond to basins of attraction around the finite number of final
configurations of atoms. The minimized energy per unit length
(equation~\ref{eqn:GrainBoundaryEnergy}) found by this method is
what we record as the grain boundary energy.

% original gimp file located in /work/thesis/writeup/grain_boundary_figures

%Data for this plot is located on ccmr in
%ASP/MDWebServices/GrainBreaker/SymmetricGrainBoundaries/SymmetricUnder20Unique/geo106

We have also tried thermal annealing and have found that for certain geometries,
the grain boundary migrates to form a jagged interface with segments of
different grain boundaries that collectively have a lower total energy than the
geometry given by the original set of tilt angles.  The tendency of certain
grain boundaries to corrugate is also discussed by Ishida and Pumphrey
\cite{bubble-raft, pumphrey}.  As a corrugated grain boundary is torn apart to
measure the cohesive law, the corners form stress concentrations which weaken
the grain boundary.  Despite the fact that this configuration may be more
natural, it is not what we intend to measure.  We wish to measure the fracture
toughness of all possible pairs of tilt angles, even if those pairs of tilt
angles happen to be unstable. In a few cases, the minimization procedure
described above produced a curved grain boundary. In these cases, we
constrained the displacements in the $y$-direction such that a consistent
pattern of flaws along a straight grain boundary is achieved.

 After finding the ideal $y$-displacement, we increment the strain by displacing
the constrained layers of atoms in the $x$-direction, away from the grain
boundary.  We relax the atoms and measure the force in the $x$-direction per
unit length on the constrained layer of atoms.  If the measurement of the stress
drops abruptly during one strain step, the simulation restores the positions
from a previous step, reduces the size of the strain increment, and proceeds.
The result of such a simulation is shown in figure~\ref{fig:CohesiveLaw}.  The
maximum stress in the stress strain curve is what we define as the peak stress.
%similar methods are used by~\cite{spearot-atomistic, shenderova-diamond}.

\section{Grain Boundary Geometries}

\subsection{Lattice Symmetries and Tilt Angles}
\label{sec:AngleRange}
\begin{figure}[thb]
\begin{center}
\includegraphics[width=\figWidth]{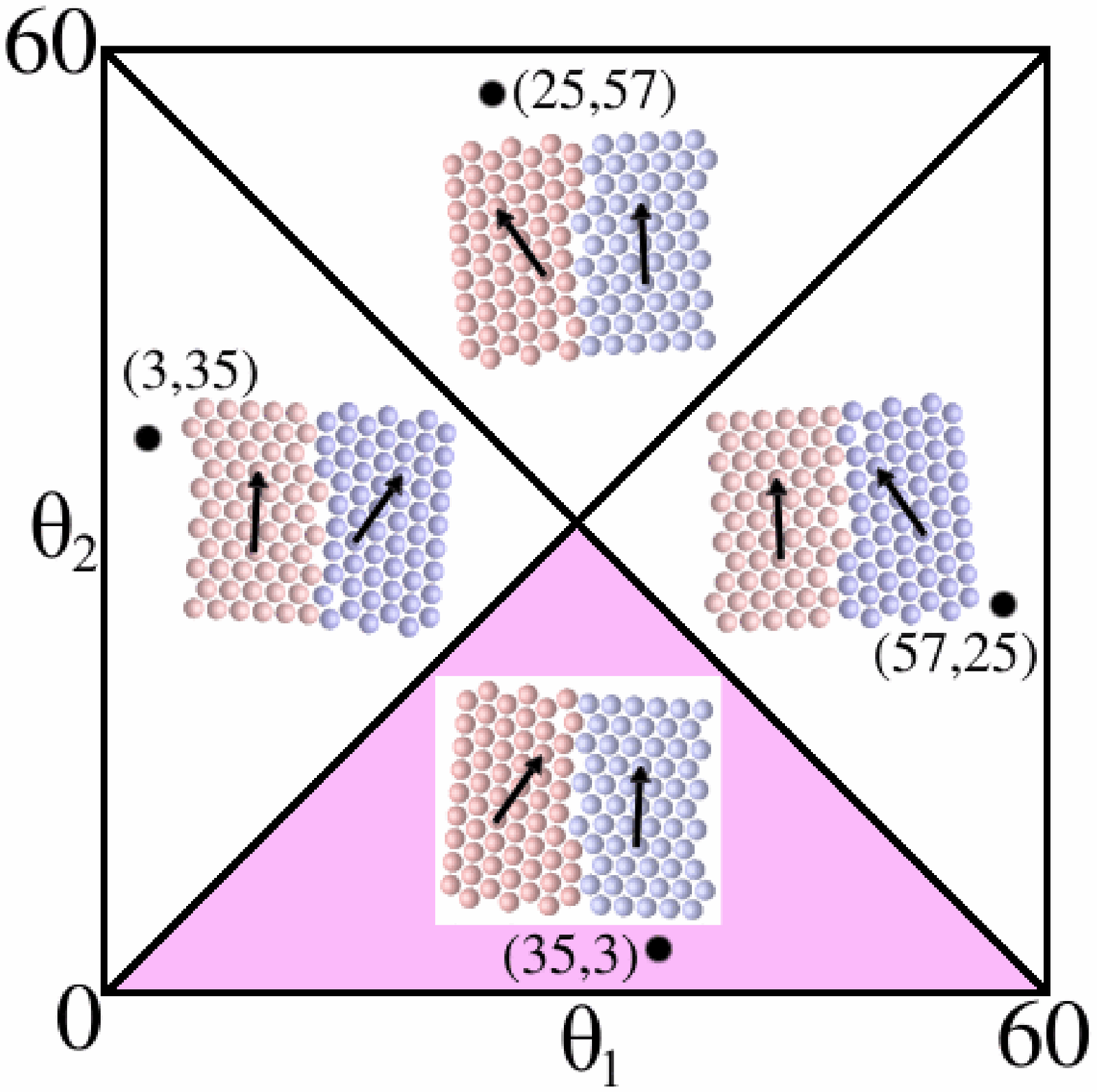}
\end{center}
\mycaption[Finding the Minimum Necessary Range of Angles.]{\coloronline For each point in the
shaded region, there is an equivalent grain boundary in each of the other
regions.}
\label{fig:AngleRange}
\end{figure}

For the triangular lattice, it is clear that we only need to explore tilt angles
between 0$^{\circ}$ and 60$^{\circ}$, but we can further reduce the space of
grain boundary geometries to consider.  Figure~\ref{fig:AngleRange} represents
the space of tilt angle pairs.  Reflecting a point in this space through the
$\theta_2 = \theta_1$ line corresponds to swapping the two grain orientations
which, as shown in figure~\ref{fig:AngleRange}, is equivalent to flipping the
grain boundary in both the vertical and horizontal directions, resulting in the
same grain boundary.  Reflecting through the $(30,30)$ point takes
$(\theta_1,\theta_2)$ to $(60-\theta_1,60-\theta_2)$, reversing the sense of
rotation of each grain.  As shown in figure~\ref{fig:AngleRange}, this is
equivalent to flipping the grain boundary in the vertical direction.  Reflecting
through the $\theta_2 = 60-\theta_1$ line takes $(\theta_1,\theta_2)$ to
$(60-\theta_2,60-\theta_1)$, both switching the grains and reversing the senses
of rotation.  This is equivalent to flipping the grain boundary in the
horizontal direction, also shown in figure~\ref{fig:AngleRange}.  Thus, we only
need to consider the pairs of tilt angles in the triangle enclosed by the
$\theta_2 = 0$ line, the $\theta_2 = \theta_1$ line, and the $\theta_2 =
60-\theta_1$ line -- the shaded region in figure~\ref{fig:AngleRange}.

% the original gimp file is in /home/work/thesis/grain_boundary_figures

\subsection{Finding All Possible Geometries for Periodic Boundary Conditions}
\label{sec:AllPossibleGeometries}

In order to better simulate the grain boundary in the bulk as described in
section~\ref{sec:measuring_cohesive_laws}, we would like to use periodic boundary conditions
along the direction of the grain boundary ($y$-direction).  
%(In the
%$x$-direction, a boundary layer of fixed atoms is used to impose the boundary
%conditions.)  
In order to simulate 2D grain boundaries in periodic boundary
conditions, we need not only for both grain orientations to have finite repeat
distances, but also for the repeat distances to be commensurate with one
another. Let $(a,b)$ be the lattice vector that is parallel to the edge of a 2D
triangular lattice.  Since the basis vectors are at a 60$^{\circ}$ angle to one
another, the repeat distance of a particular orientation of a 2D triangular
lattice is given by
\begin{equation}
D=\sqrt{(a+\frac{1}{2}b)^2+(\frac{\sqrt{3}}{2}b)^2}=\sqrt{a^2+b^2+ab}.
\label{repeatDistance}
\end{equation}
The tilt angle is then given by
\begin{equation}
\theta = \sin^{-1} \left( \frac{a+b/2}{\sqrt{a^2 +b^2 +ab}} \right).
\label{tiltangle}
\end{equation}

In order to expand the number of possible geometries, we have also considered
geometries for which applying a small strain to each grain allows us to fit both
grains inside the same periodic box.  For any pair of orientations that have
repeat distances that are not commensurate, we can find a continued fraction
approximation to the ratios of their lengths, $p/q \approx D_1/D_2$.  We can
strain each grain equally into a box of size $L=(1/2)(p D_2 + q D_1)$, where the
strain required is $|L-p D_2|/L$.  For a strain of 0.05\%, the grain boundary
would have to be 1,000 $\mbox{\AA}$ long before the strain would alter the structure
of the grain boundary.  We find all pairs of tilt angles that correspond to
commensurate or near commensurate grain boundaries by looping over pairs of
surface vectors and comparing the repeat distances.

% data and plot located in 
% /afs/msc.cornell.edu/home/jsethna/val/ASP/MDWebServices/GrainBreaker/2DGeometries
\begin{figure}
\begin{center}
\inThesis{
\includegraphics[width=12cm]{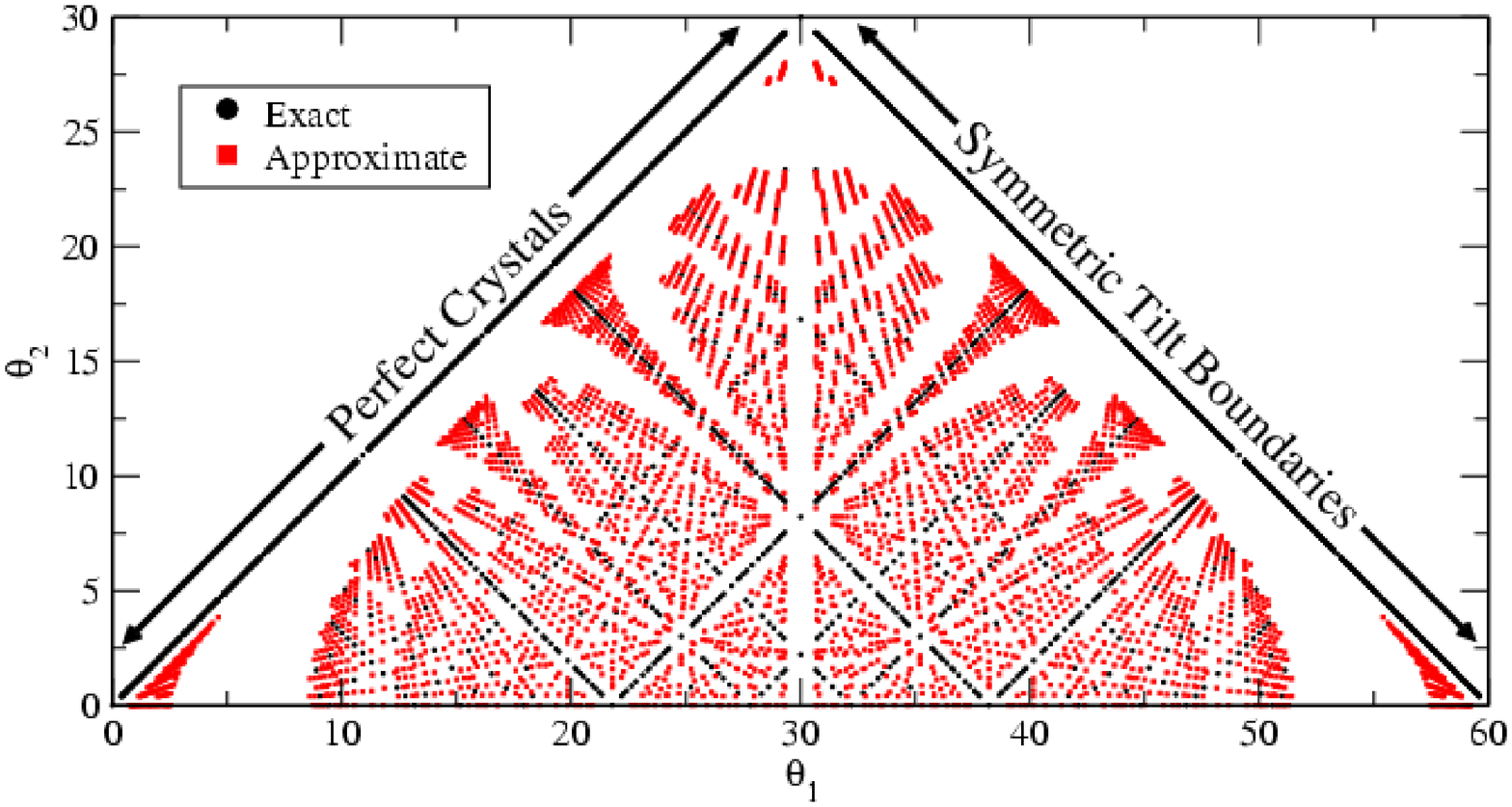}}
\inPaper{
\includegraphics[width=8cm]{grain_boundary_figures/GeometriesAnnotated}}
\end{center}
\mycaption[All Possible Geometries.]{ The set of points above represent all 2D
  grain boundary geometries that can be simulated in a periodic box of 70
  lattice constants or less, with a strain of 0.05\% or less. There are gaps near
  perfect crystals, symmetric grain boundaries, and high symmetry grain
  boundaries (discussed in section~\ref{sec:HighSymmetryGrainBoundaries})
  because creating a new, nearby geometry requires adding flaws at large
  distances.  Lines radiating from high symmetry geometries represent flaws
  added at closer and closer distances as you move away from the point
  representing the original geometry. Lines of slope $\pm 1$ represent a
  constant misorientation between the two grains.}
\label{fig:funkyplot}
\end{figure}

Figure~\ref{fig:funkyplot} shows all possible geometries that can be simulated
with a periodic length of 70 lattice constants or less, and a strain of 0.05\% or
less.  The $\theta_1=\theta_2$ line corresponds to perfect crystals, while the
$\theta_1=60-\theta_2$ line corresponds to symmetric grain boundaries.  There is
a gap near each of these lines because creating very small angle grain
boundaries, or a geometry very close to symmetric grain boundaries, requires
adding flaws that are separated by large distances.  We see gaps near other high
symmetry grain boundaries for the same reason.  High symmetry grain boundaries
are discussed in section~\ref{sec:HighSymmetryGrainBoundaries}. The lines
radiating out from high symmetry grain boundaries represent adding single flaws
to those high symmetry grain boundaries at larger and larger distances, as you
approach the high symmetry boundary.  Each line represents a different type of
flaw.  The symmetry about the $\theta_1 = 30$ line is due to the fact that
grains with tilt angles $\theta$ and $60-\theta$ have the same repeat distance.
Therefore, the grain boundaries given by the tilt angles $(\theta_1, \theta_2)$
and $(60-\theta_1,\theta_2)$ have the same overall repeat distance, though these
are different grain boundaries.

\subsection{High Symmetry Grain Boundaries}
\label{sec:HighSymmetryGrainBoundaries}
\begin{figure}
\begin{center}
%\includegraphics[height=5cm]{grain_boundary_figures/4-17Burger}
%\hskip .5cm
%\includegraphics[height=5cm]{grain_boundary_figures/14-53Burgers}
%\hskip .5cm
%\includegraphics[height=5cm]{grain_boundary_figures/19-49Burgers}
%\hskip .5cm
%\includegraphics[height=5cm]{grain_boundary_figures/19-46Burgers}
\includegraphics[width=6cm]{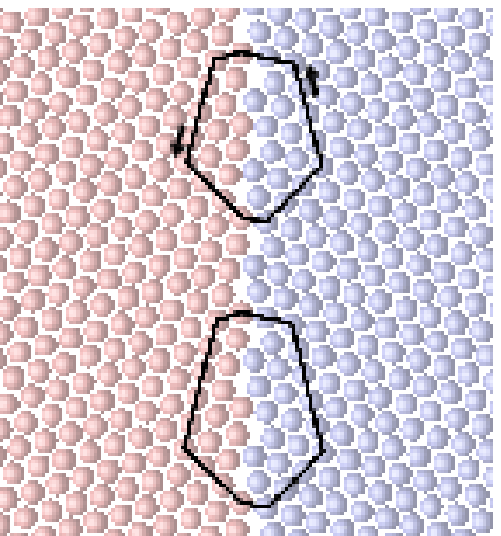}
\end{center}
\mycaption[Finding the Burger's Vector for Flaws Along Vicinal Grain
Boundaries.]{\coloronline Grain boundaries with tilt angles in the vicinity of
high symmetry grain boundaries are comprised of the pattern of structural units
of the high symmetry grain boundary with an element from the neighboring high
symmetry geometry.  When defining the Burger's vector of a flaw added to a high
symmetry grain boundary we are no longer comparing the flaw to the perfect
lattice but to the repeating pattern of structural units.  Therefore, we must
cross the grain boundary at equivalent portions of the structural unit in our
reference loop and in our loop around the flaw.  The loop on the bottom is the
reference loop around the region of the grain boundary without the added flaw.
The loop on the top surrounds the added flaw, which we model as a partial grain
boundary dislocation.  The arrows indicate the segments needed to make the top
loop match the bottom loop; for example, the arrow on the upper right indicates
that the path around the defect is one atomic length shorter on that leg than
the corresponding leg of the reference loop.  The arrows sum to the Burger's
vector of the added grain boundary dislocation. This grain boundary has a tilt
angle of 44.3$^{\circ}$ and the added partial dislocation has a Burger's vector
with a length of 0.495 lattice constants and a direction along the negative
$x$-axis.
%The first
%  grain boundary has a tilt angle of 49.67$^{\circ}$ and a Burger's vector with
%  a length of 0.38 inclined at 210$^{\circ}$ and 150$^{\circ}$ from the
%  $x$-axis.  The second grain boundary has a tilt angle of 48.58$^{\circ}$
%Name all of the
%  burger's vectors. describe transitioning between geometries, how to find
%  burger's vector...
}
\label{fig:AddingFlaws}
\end{figure}

Certain grain boundary geometries have particularly low repeat distances.  These
grain boundaries have special properties.  They mark the center of cusps in the
grain boundary energy and discontinuous increases in the fracture strength as a
function of tilt angle.  Table~\ref{table:HighSymmetry} show examples of high
symmetry grain boundaries and figure~\ref{fig:AddingFlaws} shows examples of
geometries in between, which constitute adding a single flaw per supercell
repeat distance to the high
symmetry grain boundary.  Note how the added flaws constitute a compromise
between the two high symmetry geometries.  Furthermore, a less high symmetry
grain boundary can be repeated, and again have a single flaw added.  Because of
this hierarchical procedure for constructing lower-symmetry boundaries,
the grain boundary energy and fracture strength as a function of tilt angle will
have a self similar nature.

\begin{table}[thb]
\begin{center}
\caption[High Symmetry Grain Boundary Geometries]{ {\bf High Symmetry Grain Boundary Geometries}}
\label{table:HighSymmetry}
\begin{tabular}{|c|c|c|c|c|}
\inThesis{
\hline $\theta$ & Miller Indices & Repeat Distance (lattice constants) &
$\Sigma$ & Structure \\}
\inPaper{
\hline $\theta$ & Miller & Repeat Distance &
$\Sigma$ & Structure \\
& Indices & (lattice constants) & & \\}

\hline
49.10 & (1,4) & 2.64 & 7 &
\includegraphics[height=1cm]{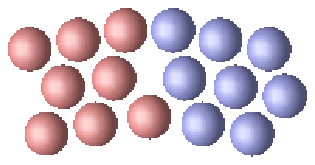}
\\

\hline
43.89 & (2,5) & 3.61 & 13 &
\includegraphics[height=1cm]{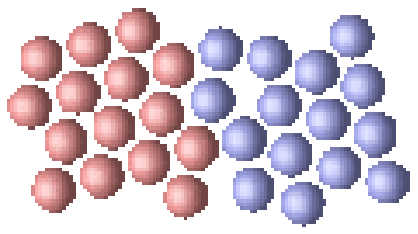}\\

\hline 53.41 & (1,7) & 4.35 & 19 &
\includegraphics[height=1cm]{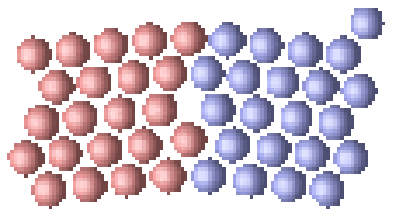}\\

\hline  40.89 & (1,2) & 4.58 & 21 &
\includegraphics[height=1cm]{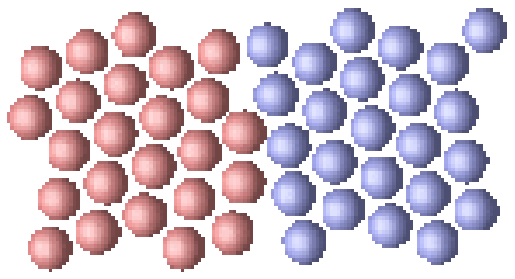}\\

\hline  38.94 & (4,7) & 5.56 & 31 &
\includegraphics[height=1cm]{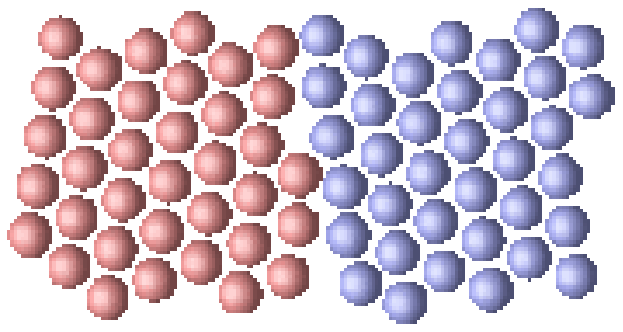}\\

\hline  51.78 & (2,11) & 7.0 & 49 &
\includegraphics[height=1cm]{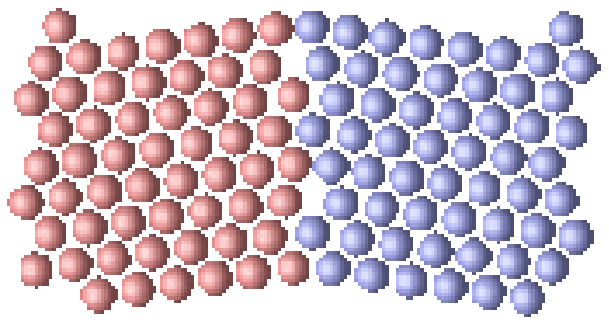}\\

\hline
\end{tabular}
\end{center}
\end{table}

The coincidence site lattice model (CSL) describes grain boundaries in terms of
$\Sigma$, the inverse density of lattice sites that are shared by the two grain
orientations when rotated about a common lattice point~\cite{bollmann}.  In 2D,
CSL grain boundaries are necessarily commensurate and vice versa.  For the
triangular lattice, each common lattice point for a given pair of grain
orientations has a repeat cell that is an equilateral triangle with one edge
defined by the surface vector with length $D$.  $\Sigma$ is equal to the number
of lattice points inside this cell and is given by the area of the cell divided
by the area of one lattice triangle, $\Sigma = D^2$.  While the CSL formulation
gives a simple method for finding commensurate grain boundaries, it is
misleading to suggest that the actual coincidence of sites plays a physical
role.  In fact, shifting the grains in the $y$-direction so as to minimize the
grain boundary energy (described in section~\ref{sec:measuring_cohesive_laws}) generally causes
the atoms to no longer coincide.  Without this shift, the two free surfaces will
have atomic planes that meet at the same point.  The elastic energy is lowered
by staggering the dislocations~\cite{HirthAndLothe}.  Wolf has pointed out that
for 3D, the CSL formulation of grain boundaries involves a redundant number of
parameters~\cite{wolf-structure}.

Bishop and Chalmers introduced the concept of ``structural units'' - polygonal
structures of atoms along high angle grain boundaries~\cite{bishop-chalmers}.
We find that the patterns of ``structural units'' are more relevant to the
physical properties of the grain boundaries. When relaxed, the structural units
of grain boundaries in 2D, triangular lattices are 5 atoms forming a regular
pentagon or a pentagon that is slightly stretched.  Each high symmetry grain
boundary is comprised of a simple pattern of pentagonal structural units (table
\ref{table:HighSymmetry}).  When a high symmetry grain boundary is perturbed, an
element of the pattern of structural units from the neighboring high symmetry
grain boundary is introduced (figure~\ref{fig:AddingFlaws}).  As the tilt angles
move closer to the neighboring high symmetry grain boundary, the flaws become
closer together until they outnumber the original structure.  The roles of flaw
and original structure are then reversed.  In this manner, combinations of
patterns of structural units can be used to build up any commensurate, high
angle grain boundary.

We find the structural units a useful way of conceptualizing the different grain
boundary geometries.  We have not found the coincidence site lattice model
useful in our investigations.  In systematically constructing grain boundaries
with small repeat distances (section~\ref{sec:AllPossibleGeometries}), 
%and appendix~\ref{sec:3D_grain_boundaries}), 
we have found the surface lattice
vectors of the two sides of the grain boundary to be a useful description.  We
suggest that the structure of high angle grain boundaries can best be described
in terms of a dislocation model.  The ``extra flaws'' added to create vicinal
grain boundaries can be described as partial dislocations.
Figure~\ref{fig:AddingFlaws} shows how to find the Burger's vectors for the
added flaws by examining the flaw in the pattern of structural units. \footnote{Since
this Burger's vector may be a sum of lattice vectors from either
side of the grain boundary, the total may not be a full translation vector of the
triangular lattice.}
We will
show in section \ref{sec:HighAnglePeakStress} that this model gives excellent
agreement in the stress field due to the added flaws, outside the background of
the original flaw structure.  We will show in
sections~\ref{sec:grain_boundary_energy} and \ref{sec:fracture_strength} that
the dislocation model provides the most powerful framework for understanding the
geometry dependence of the properties of grain boundaries.

\begin{comment}

\section{Measuring the Cohesive Law}

The most natural grain boundary configuration is found using the method
described in section~\ref{sec:Minimize}. 

At high angle grain boundaries, the fracture is very brittle in nature.  With a
wide enough simulation, as discussed in section~\ref{sec:Width}, the grain
boundary snaps open shortly after reaching the peak stress.  For lower angle
grain boundaries, the failure is not always brittle. Symmetric, low angle grain
boundaries with tilt angles near 0$^{\circ}$ undergo intragranular fracture,
with the crack nucleating at a single dislocation and growing at an angle
30$^{\circ}$ from the grain boundary direction
(figure~\ref{fig:LowAngleIntragranularFracture}). Symmetric grain boundaries
with tilt angles near 30$^{\circ}$ fail with alternate dislocations gliding in
opposite directions unless the glide is prevented by the boundaries of the
narrow simulation (figure~\ref{fig:LowAngleIntergranularFracture}).  In this
case, the grain boundaries undergo brittle fracture with a predictable peak
stress.  This is discussed further in section \ref{sec:LowAngleGrainBoundaries}

Sections~\ref{sec:AngleRange} and~\ref{sec:AllPossibleGeometries} explain the
minimum range of pairs of tilt angles necessary to explore all geometries and
how to find the geometries that can be simulated in periodic boundary
conditions. 

%[Asymmetric grain boundaries? Systematic exploration of size?
%  Explore line radiating from high symmetry boundary?]

\end{comment}

\section{Grain Boundary Energy}
\label{sec:grain_boundary_energy}

We have measured the grain boundary energy and peak stress for all symmetric
grain boundaries with repeat distances under 20 lattice constants and asymmetric
grain boundaries with repeat distances under 30 lattice constants.  In order to
explore the regions close to high symmetry grain boundaries, we have added a few
geometries with longer repeat distances, close to the high symmetry grain
boundaries.  

\begin{figure}[thb]
\begin{center}
\inThesis{
\includegraphics[width=12cm]{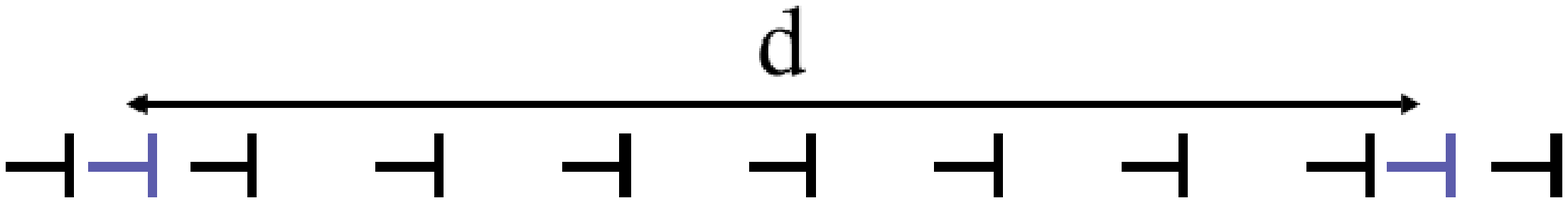}}
\inPaper{
\includegraphics[width=8cm]{grain_boundary_figures/ExtraDislocations}}
\end{center}
\mycaption[Adding a Flaw to a High Symmetry Grain Boundary.]{\coloronline The lighter
  dislocations represent flaws added a distance $d$ apart, to an existing
  pattern of dislocations, shown in black, with a short repeat distance.  The
  added flaws can also move, screen, or cancel the flaws that make up the high
  symmetry boundary.}
\label{fig:ExtraDislocations}
\end{figure}

The energy~\footnote{We use 'energy' for 'energy per unit length' throughout.}
associated with the series of flaws that make up the grain boundary is defined
in equation~\ref{eqn:GrainBoundaryEnergy}.  As stated earlier, and found by
several earlier
studies~\cite{sansoz-molinari-structure,chen-GB,old-grain-boundary-sims,pumphrey,
wolf-structure,palumbo-aust,wolf-jaszczak}, cusps appear at high symmetry
boundaries.  For small angle grain boundaries, the grain boundary energy has the
form
\begin{equation}
E_{GB} = \frac{ \mu b}{4 \pi (1-\nu)} | \theta| \log \left( \frac{e \alpha }{2 \pi
|\theta|} \right)  \propto |\theta| \log |\theta|.
\label{SmallAngleGrainBoundaryEnergy}
\end{equation}
where $\mu$ is the shear modulus, $b$ is the Burger's vector, $\nu$ is the
Poisson ratio, and $\alpha$ is a factor that includes the core energy
\cite{HirthAndLothe}.  One can now imagine the same scenario applied to high
symmetry boundaries.  We can take a high symmetry grain boundary and add or
subtract flaws a distance $d$ apart as shown in figure
\ref{fig:ExtraDislocations}.  By the same reasoning as used for the low angle
grain boundaries, the energy near the high symmetry grain boundary will have the
form
\begin{equation}
E_{GB} = E_0 + \frac{ \mu b}{4 \pi (1-\nu)} |\theta - \theta_0| \log \left(
\frac{e \alpha'}{2 \pi |\theta - \theta_0|}\right)
\label{eqn:HighSymmetryGrainBoundaryEnergy}
\end{equation}
where $E_0$ is the energy of the high symmetry grain boundary which occurs at
the angle $\theta_0$ and $\alpha'$ incorporates the core energy of the flaw within
the pattern of flaws.  

% the original figure is in /home/val/thesis/writeup/grain_boundary_geometries

% the data for this plot is in the ccmr directory
% ASP/MDWebServices/GrainBreaker/SymmetricGrainBoundaries
\begin{figure}[thb]
\begin{center}
\includegraphics[width=\figWidth]{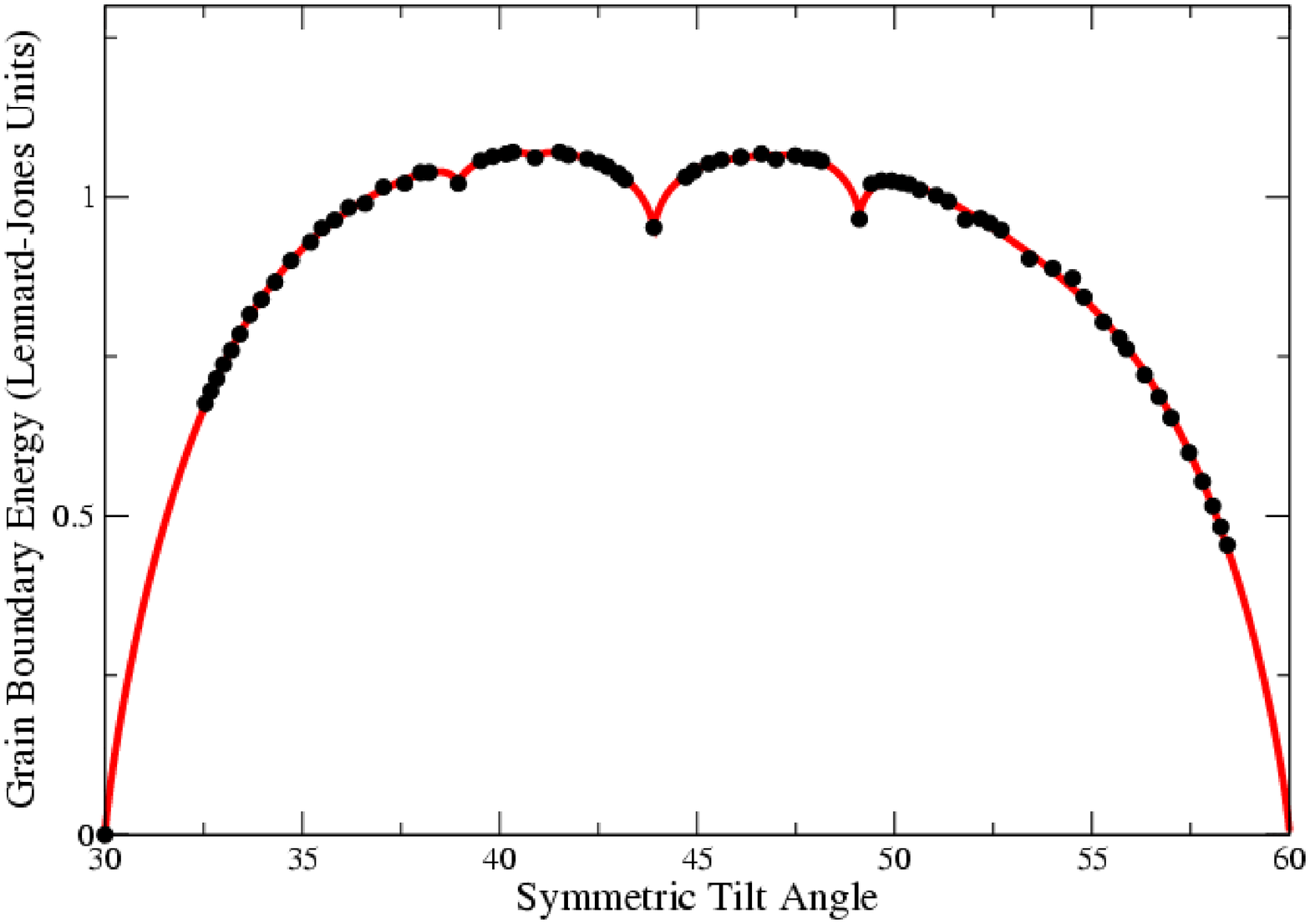}
\end{center}
\mycaption[ Grain Boundary Energies for Symmetric Geometries.]{\coloronline Cusps appear at
  high symmetry grain boundaries (listed in table~\ref{table:HighSymmetry}) and
  have the same $\theta \log \theta$ shape as the energy of low angle grain
  boundaries. The line is the fit given by
  equation~\ref{eqn:GrainBoundaryEnergyFittingFunction}. Notice that hints of
  smaller cusps are visible in the data.}
\label{fig:GrainBoundaryEnergy}
\end{figure}

The grain boundary energies for all of the symmetric grain boundaries that we
have measured are shown in figure~\ref{fig:GrainBoundaryEnergy}.  Note that
cusps occur at the angles listed in table~\ref{table:HighSymmetry}.  We are able to
fit the data for symmetric grain boundaries to a function of the form

\begin{eqnarray*}
E_{GB}(\theta) 
&=& a_0 |\sin 3\theta| \log \frac{b_0}{|\sin 3 \theta|}
+ 
\inThesis{+} \inPaper{\\ &+&} a_{30} |\sin 3(\theta-30)| \log \frac{b_{30} }{|\sin 3( \theta -30)|} \nonumber \\
&+& \sum_{i=0}^n \Bigg( a^{(s)}_i | \cos 6 \theta 
- \cos 6 \theta_i | 
\inPaper{\\ & &}
\log \frac{ b^{(s)}_i }
{ |\cos 6 \theta - \cos 6 \theta_i | } \nonumber \\
&+&  a^{(a)}_i ( \cos 6 \theta - \cos 6 \theta_i )
\inPaper{\\ & &}
\log \frac{b^{(a)}_i }{ | \cos 6 \theta - \cos 6 \theta_i |} \Bigg) \nonumber \\
&+& \sum_{j=0}^m c_j \cos(6 j \theta) + d 
\label{eqn:GrainBoundaryEnergyFittingFunction}
\end{eqnarray*}
The first two terms fit the cusps at 0$^{\circ}$ and 30$^{\circ}$, where the
cusps are symmetrical about their respective center points.  The next set of
terms in the sum fit the cusps at high symmetry tilt angles.  The function $|
\cos 6 \theta - \cos 6 \theta_i |\log \frac{1 }{ | \cos 6 \theta - \cos 6
\theta_i |}$ was chosen as a fitting function because it asymptotically gives a
$\theta \log (1/\theta)$ shaped cusp in the near vicinity of $\theta_i$ and
because it has the correct symmetry: even mirror symmetry at 0$^{\circ}$ and
30$^{\circ}$ and an overall period of 60$^{\circ}$.  We use one term that is
antisymmetric about $\theta_i$ and one term that is symmetric about $\theta_i$
so that we can fit the shape on either side of the cusp independently.  We do
not expect the slope of the curve on either side of the cusp to be the same
since the Burger's vectors of the additional flaws for the geometries on either
side of the high symmetry geometry may differ (figure~\ref{fig:AddingFlaws}).

The $\theta_i$ can be any angles that have the shortest repeat distances, such
as those given in table~\ref{table:HighSymmetry}.  The curve in figure
\ref{fig:GrainBoundaryEnergy} is the result of fitting equation
\ref{eqn:GrainBoundaryEnergyFittingFunction} to the data shown in the same
figure.  We have used $\theta_i = (49.10^{\circ}, 43.89^{\circ}, 40.89^{\circ},
38.94^{\circ})$ since these angles have particularly prominent cusps and short
Miller indices (given in table \ref{table:HighSymmetry}).  Three smooth
sinusoidal terms were used in the final sum over $\cos 6 j \theta$.

The result is analogous to a devil's staircase, with a cusp singularity at each
angle that corresponds to a special rational number. In principle, the energy
has a logarithmic cusp at a dense set of points but (as is typical for devil's
staircases) the high-order cusps rapidly diminish in size.

\begin{comment}
{c[1] -> -0.140239, c[2] -> 1.70895, c[3] -> -0.0697545,
aa[3] -> -0.00074014, aa[4] -> -0.0235427, aa[5] -> -0.00891309, aa[6] -> 0.0395999, 
ba[3] -> 2.60614, ba[4] -> -13.9382, ba[5] -> 1.56146, ba[6] -> -5.53842,
as[1] -> 1.83278, as[2] -> 1.64819, 
as[3] -> 0.765246, as[4] -> 0.681113, as[5] -> 0.217407, as[6] -> 0.443717, 
bs[1] -> 0.542197, bs[2] -> 1.42121, 
bs[3] -> -1.62837, bs[4] -> -1.22911, bs[5] -> -2.26499, bs[6] -> -1.8754, 
d -> 0.0939106}
1 = 0 angle
2= 30 angle
\end{comment}

\begin{comment} % The version of the table where b is not in the log
\hline $i$ & $a^{(s)}_i$ & $a^{(a)}_i$ & $b^{(s)}_i$ & $b^{(a)}_i$ \\
\hline 1 & 0.76 & $7.4 \times 10^{-4}$ & -1.63 & 2.61 \\
\hline 2 & 0.68 & -0.023         & -1.23 & -13.9 \\
\hline 3 & 0.22 & $-8.9 \times 10^{-3}$ & -2.26 & 1.56 \\
\hline 4 & 0.44 & 0.039          & -1.87 & -5.54 \\

\hline \hline
  & $a_0$ & $ a_{30}$ & $b_0$ & $b_{30}$ \\
\hline  & 1.83 & 1.65  & 0.54 & 1.42 \\

\hline \hline
  & $c_1$ & $c_2$ & $c_3$ & $d$ \\
\hline  & -0.14 & 1.71 & -0.070 & 0.094 \\
\end{comment}

\begin{table}[thb]
\begin{center}
\caption[Coefficients for Fitting Grain Boundary Energy to
    Eq.~\ref{eqn:GrainBoundaryEnergyFittingFunction}.]{ {\bf Coefficients for Fitting Grain Boundary Energy to
    Eq.~\ref{eqn:GrainBoundaryEnergyFittingFunction}.}}
\label{table:Coefficients}
\begin{tabular}{|c|c|c|c|c|}

\hline $i$ & $a^{(s)}_i$ & $a^{(a)}_i$ & $b^{(s)}_i$ & $b^{(a)}_i$ \\
\hline 1 & 0.76 & $7.4 \times 10^{-4}$ & 0.196 & 13.6 \\
\hline 2 & 0.68 & -0.023         & 0.292 & $9.19 \times 10^{-7}$ \\
\hline 3 & 0.22 & $-8.9 \times 10^{-3}$ & 0.104 & 4.76 \\
\hline 4 & 0.44 & 0.039          & 0.154 & $3.93 \times 10^{-3}$ \\

\hline \hline
  & $a_0$ & $ a_{30}$ & $b_0$ & $b_{30}$ \\
\hline  & 1.83 & 1.65  & 1.72 & 4.14 \\

\hline \hline
  & $c_1$ & $c_2$ & $c_3$ & $d$ \\
\hline  & -0.14 & 1.71 & -0.070 & 0.094 \\

\hline
\end{tabular}
\end{center}
\end{table}

\begin{figure}[thb]
\begin{center}
\includegraphics[width=8.5cm]{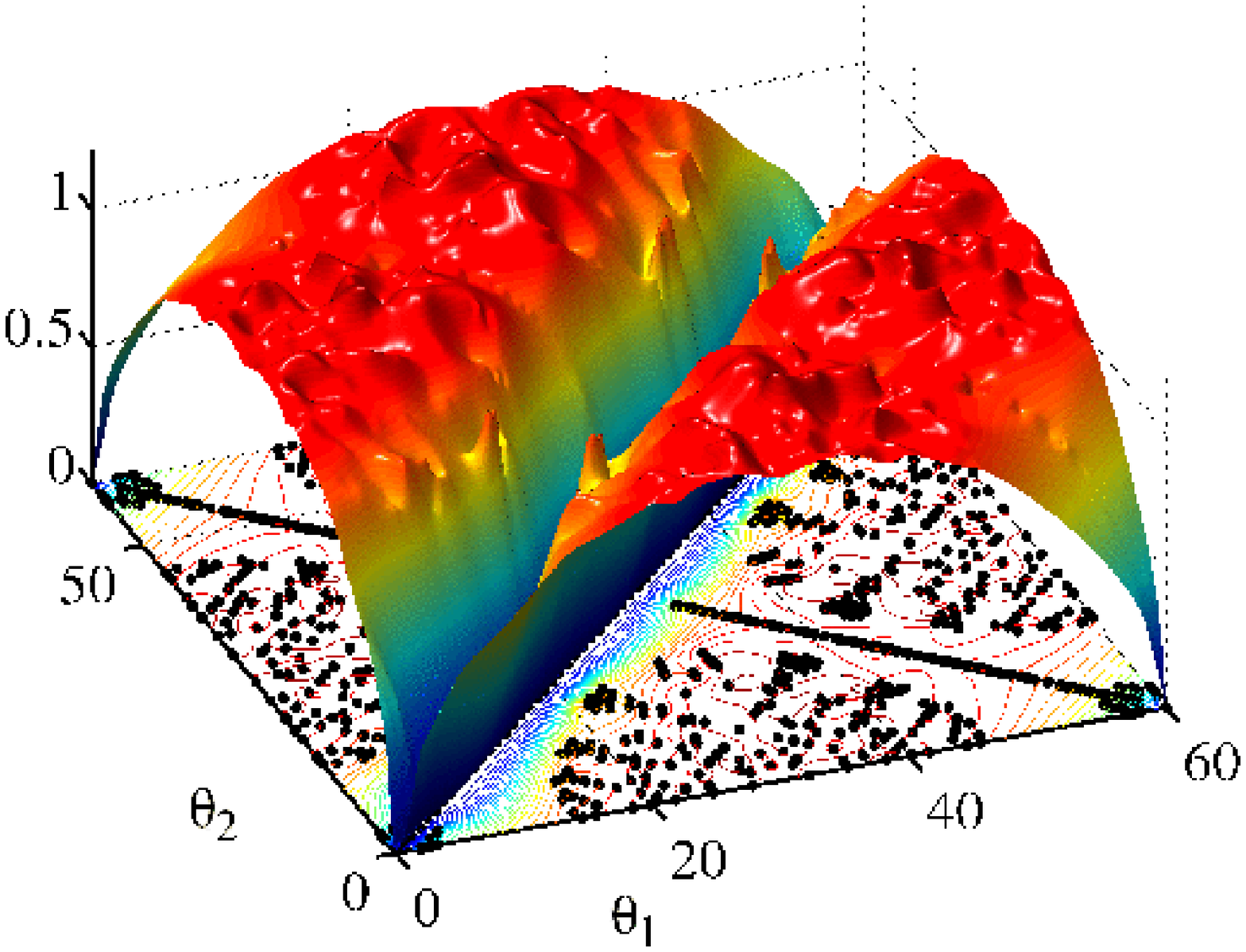}
\mycaption[Grain Boundary Energies for Asymmetric Geometries.]{The surface and
contour represents a smooth interpolation between the particular geometries that
have been simulated, represented by the black dots in the $\theta_1\theta_2$
plane.  Perfect crystals have zero grain boundary energy, shown by the groove along the
$\theta_1=\theta_2$ line.}
\label{fig:AsymmetricEnergies}
\end{center}
\end{figure}

\begin{figure}[thb]
\begin{center}
\includegraphics[width=8.5cm]{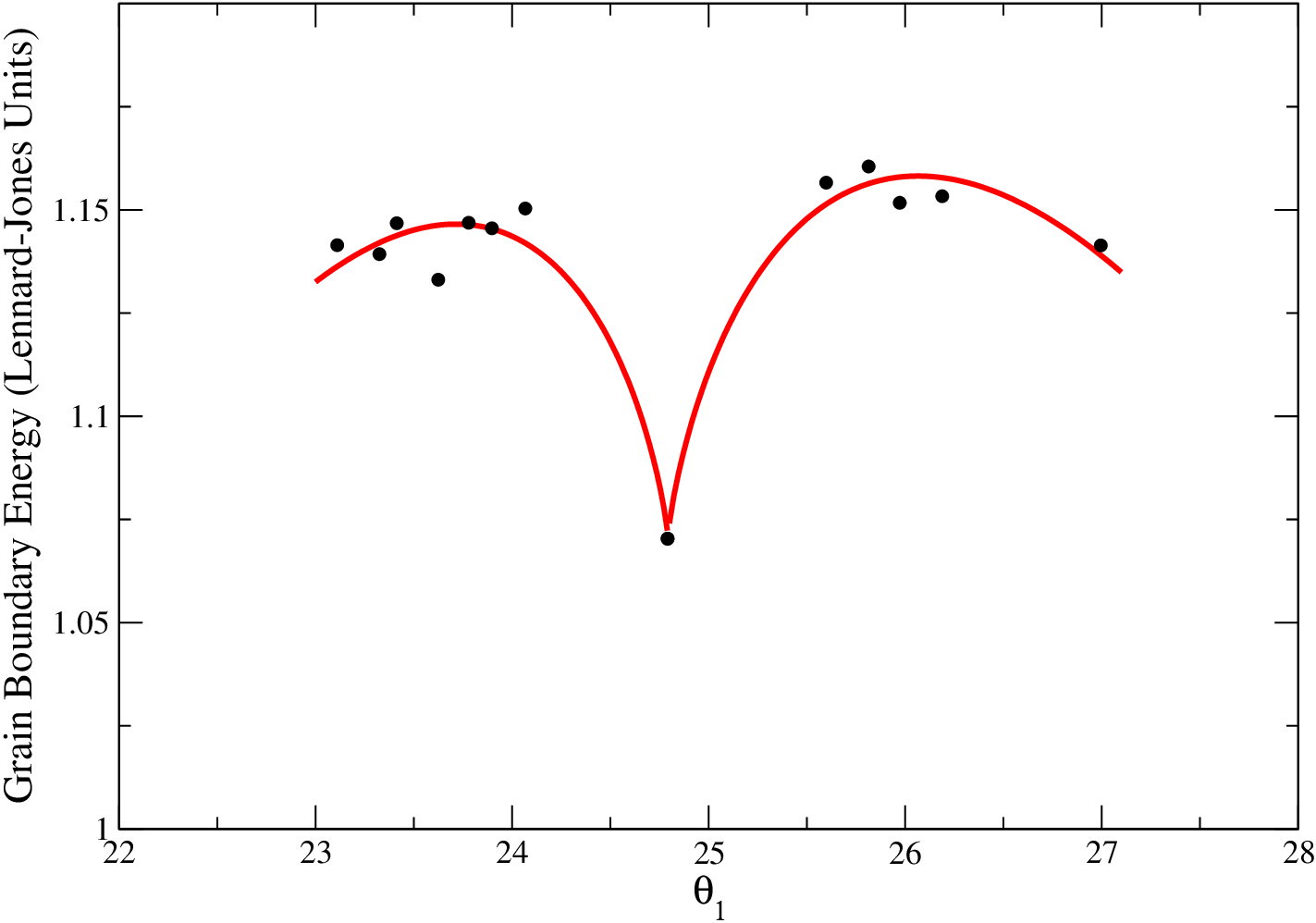}
\mycaption[Asymmetric Grain Boundary Energies in Neighborhood of High Symmetry
Geometry.]{The grain boundary geometries considered in the above plot each have
a misorientation of 21.79$^\circ$ and are in the vicinity of the high symmetry
grain boundary with tilt angles 24.79 and 3.00.  The lines are fit to
$\theta\log\theta$ for the points on either side of the cusp.} 
\label{fig:AsymmetricEnergiesNeighborhood}
\end{center}
\end{figure}

The results for the energy and peak stress of all of the grain boundaries we
have simulated are shown in figure~\ref{fig:AsymmetricEnergies}.  As with the
symmetric grain boundaries, we see energy cusps for grain boundaries with
particularly low repeat distances.  For example, one of the cusps in
figure~\ref{fig:AsymmetricEnergies} is centered around the grain boundary with
tilt angles 24.79$^\circ$ and 3.00$^\circ$ which has a repeat distance of 9.53 lattice
constants. Figure~\ref{fig:AsymmetricEnergiesNeighborhood} shows geometries in
the neighborhood of this high symmetry boundary with a constant misorientation.
We see a similar, $\theta\log\theta$ shape to the cusp centered around the high
symmetry geometry, suggesting that the same concept of adding flaws to the
short-repeat distance grain boundary applies to asymmetric 2D grain boundaries
along constant misorientation lines.

\section{Fracture Strength}
\label{sec:fracture_strength}

\subsection{Low Angle Symmetric Grain Boundaries}
\label{sec:LowAngleGrainBoundaries}
\begin{figure}[thb]
\begin{center}
\inThesis{
\includegraphics[width=5cm]{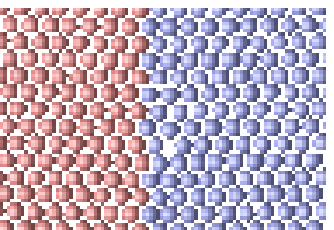}
\hskip .5cm
\includegraphics[width=5cm]{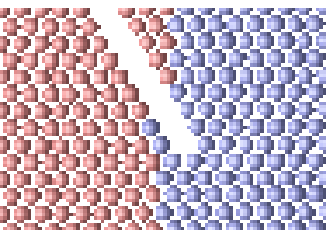}}
\inPaper{
\includegraphics[width=3.75cm]{grain_boundary_figures/SingleDislocationPerRepeatFrame01Cropped}
\hskip .5cm
\includegraphics[width=3.75cm]{grain_boundary_figures/SingleDislocationPerRepeatFrame39Cropped}}
\mycaption[Intragranular Fracture for Low Angle Grain Boundaries.]{\coloronline The figure on
  the left shows a grain boundary with symmetric tilt angle 0.81$^{\circ}$ at 0
  strain.  The same grain boundary is shown on the right with a strain of
  3.125\%. Symmetric low angle grain boundaries centered around the 0$^{\circ}$
  orientation (Miller indices $ (0,1)$) fail via intragranular fracture rather
  than intergranular fracture.}
\label{fig:LowAngleIntragranularFracture}
\end{center}
\end{figure}

The Frank conditions state that the total Burger's vector for the dislocations
making up a low angle grain boundary is equal to the difference of the surface
vectors that define the orientation of each grain.  In order to guarantee that
the grain boundary will have only one dislocation per repeat distance   after it is
relaxed, we must choose surface vectors that have a difference equal to a basis
vector and have the same repeat distance.  One pair of surface vectors
%the only other possibilities are equivalent
is $(2n+1, -n)$ and $(2n+1,-n-1)$.  This gives a symmetric grain boundary with a
single dislocation with tilt angles close to 0$^{\circ}$, a Burger's vector equal
to $(0,1)$ and a repeat distance of $\sqrt{3n^2 +3n +1}$.  Our simulations show
that such grain boundaries fail via intragranular fracture rather than
intergranular fracture as shown in figure
\ref{fig:LowAngleIntragranularFracture}.

%/home/val/work/DataStore/SingleDislocation/SingleDislocationPerRepeat-W/geo10003/

%We have also explored low angle grain boundaries centered around the
%30$^{\circ}$ lattice orientation. 
The apex of the triangle $\theta_1=\theta_2=30^{\circ}$ in
figure~\ref{fig:funkyplot} is a perfect crystal oriented at 30$^{\circ}$, which
is the end point of a series of symmetric tilt boundaries.
At exactly $\theta_1=\theta_2=30^{\circ}$, there is an abrupt
jump up in fracture strength since the perfect crystal has no ``extra
dislocations'' at which to nucleate fracture.  Our simulations find a peak
stress of 4.31 (Lennard-Jones Units) for the perfect crystal.  

\begin{figure}[thb]
\begin{center}
\inThesis{
\includegraphics[height=5cm]{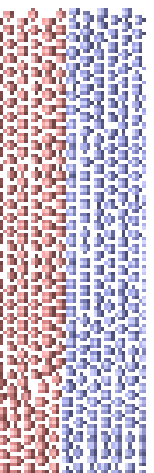}
\hskip .5cm
\includegraphics[height=5cm]{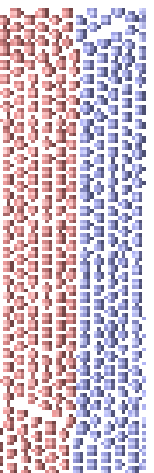}
\hskip .5cm
\includegraphics[height=5cm]{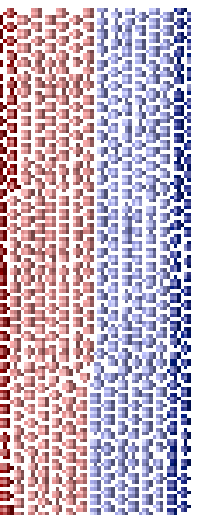}
\hskip .5cm
\includegraphics[height=5cm]{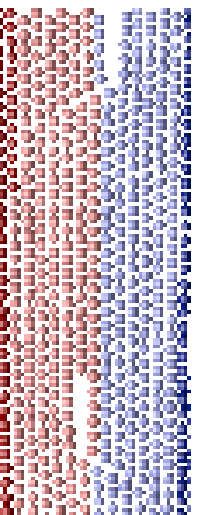}}
\inPaper{
\includegraphics[height=4cm]{grain_boundary_figures/SymmetricLowAngle-WFrame01Cropped}
\hskip .5cm
\includegraphics[height=4cm]{grain_boundary_figures/SymmetricLowAngle-WFrame31Cropped}
\hskip .5cm
\includegraphics[height=4cm]{grain_boundary_figures/SymmetricLowAngleFrame01Cropped}
\hskip .5cm
\includegraphics[height=4cm]{grain_boundary_figures/SymmetricLowAngleFrame31Cropped}}
\mycaption[Intergranular Fracture for Low Angle Grain Boundaries.]{\coloronline The first
figure shows a grain boundary with symmetric tilt angle of 30.96 at 0 strain.
The second figure shows the same grain boundary at 3.625\% strain.  In this
wider simulation, the dislocations glide apart (thin diagonal white stripes).  The
third and fourth figures shows a narrower simulation of the same tilt angle
where the intergranular fracture nucleates at each dislocation.}
\label{fig:LowAngleIntergranularFracture}
\end{center}
\end{figure}

The low angle grain boundaries near the 30$^{\circ}$ lattice orientation have
surface vectors $(1,n)$ and $(n,1)$, repeat distances $\sqrt{1+n+n^2}$, and
total Burger's vector $(-1,1)$, which splits into two flaws with Burger's
vectors $(0,1)$ and $(-1,0)$ shown in figure
\ref{fig:LowAngleIntergranularFracture}.  For wide enough simulations, these
dislocations glide in opposite directions until they are restricted by the
constrained zones on either side.  For narrower simulations, the dislocations do
not glide but form nucleation points for grain boundary fracture causing an
abrupt jump down in the peak stress compared to the peak stress of the perfect
crystal.

%/home/val/work/DataStore/SingleDislocation/SymmetricLowAngle-W/geo10003/
%/home/val/work/DataStore/SingleDislocation/SymmetricLowAngle/geo10003/

%/home/val/work/DataStore/SingleDislocation/SymmetricLowAngle
\begin{figure}[thb]
\begin{center}
\includegraphics[width=\figWidth]{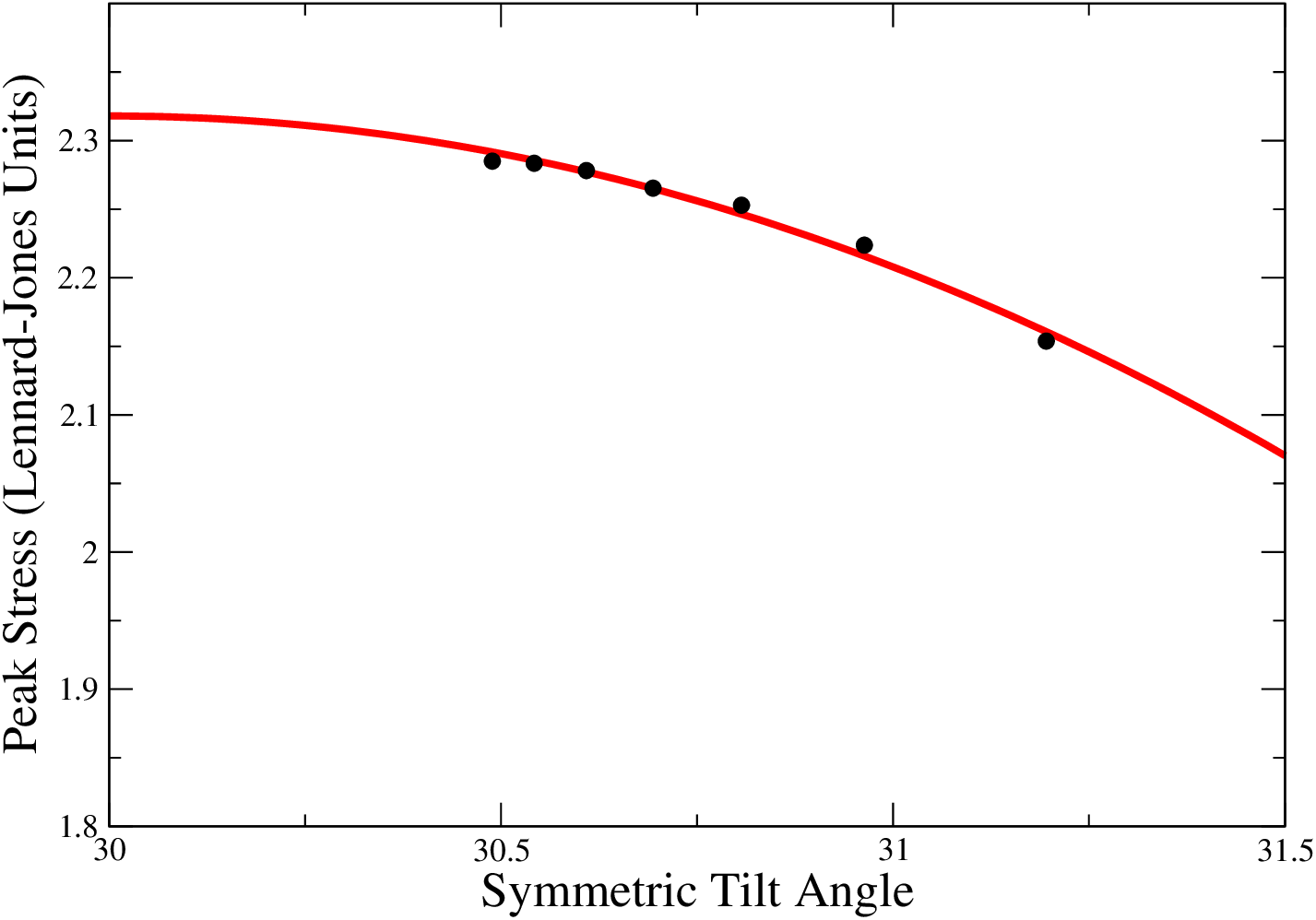}
\end{center}
\mycaption[Peak Stress vs. Tilt Angle For Low Angle Grain
Boundaries.]{\coloronline The peak
stress has a large jump downward as soon as a grain boundary deviates from
$\theta=30^{\circ}$ (perfect crystal).  Here the peak stress for the perfect
crystal, at the 30$^{\circ}$ orientation is 4.31 (Lennard-Jones units), which
would require a vertical scale 11 times as big.  The dislocations forming the
boundary act as nucleation sites for fracture no matter how far apart they are.
After this jump the peak stress has a parabolic dependence on angle for low
angle grain boundaries that are constrained by width to fracture in
qualitatively similar ways.  Fracture nucleates exactly at the dislocation and
the first non-vanishing term in the stress at this point due to neighboring
dislocations goes as $1/d^2$, where d is the distance between dislocations. }
\label{fig:LowAnglePeakStress} \end{figure}

%Let the stress at which the added dislocation nucleates %fracture be
%$\sigma_c$. 
For narrow simulations of low angle grain boundaries in this
region, we find that the peak stress has a parabolic dependence on angle, shown
in figure~\ref{fig:LowAnglePeakStress}. We can explain this parabolic dependence
as due to a partial screening of the external stress $\sigma_{ext}$ by the
neighboring dislocations on the grain boundary. Assume that the dislocation
has a critical stress for nucleating fracture equal to $\sigma_c$.  The
dislocation feels a stress due to its neighboring dislocations, each a distance
$d$ apart, in addition to the external, applied stress.  The total stress felt
by each dislocation can be written
\begin{equation}
\sigma = \sigma_{ext} + \sum_{neigh.\ disloc.} \sum_{n=0}^{\infty} \frac{a_n}{d^n}
\label{eqn:StressFeltByDislocation}
\end{equation}
where $\sigma_{ext}$ is the external stress. The $n=1$ term is the Volterra
solution given by 
\begin{eqnarray}
\sigma_{xx}(x,y) &=& - \frac{\mu b}{2 \pi (1-\nu)} \frac{y (3 x^2 +
    y^2)}{(x^2+y^2)^2} \\
\sigma_{yy}(x,y) &=&  \frac{\mu b}{2 \pi (1-\nu)} \frac{y (x^2 -
    y^2)}{(x^2+y^2)^2} \\
\sigma_{xy}(x,y) &=&  \frac{\mu b}{2 \pi (1-\nu)} \frac{x (x^2 -
    y^2)}{(x^2+y^2)^2}
\label{eqn:VolterraSolution}
\end{eqnarray}
where the $x$-direction is the direction of the Burger's vector.  Since the
Volterra solution is odd, the $n=1$ term of the stress at each dislocation
vanishes as we sum over the neighboring dislocations on either side.  The first
nonvanishing term in equation~\ref{eqn:StressFeltByDislocation} is the $n=2$
term which has three contributions. (1) The $n>1$ terms are the multiple expansions
of the stress field \cite{NicksThesis} as well as nonlinear terms. (2) The nonlinear
term in strain field has the form $du/dx * du/dx$, giving a power law of
$1/r^2$.
%, which contributes to the $n=2$ term.  Other corrections also contribute
%to the $n=2$ term.  
(3) Geometrical restrictions cause some grain boundaries to have
flaws unequally spaced in the $y$-direction (though for the results given in
figure~\ref{fig:LowAnglePeakStress} we have only explored geometries with
equally spaced flaws).  The grain boundaries geometries used in figure
\ref{fig:LowAnglePeakStress} do have flaws that are not aligned perfectly in the
$x$-direction.  In each of these cases, shifting the dislocation constitutes
adding a dislocation dipole (adding one positive and one negative, canceling a
dislocation and adding a new one), and therefore is another contribution to the
$1/r^2$ term.  The external stress needed to produce a stress equal to
$\sigma_c$ at each flaw is then
\begin{equation}
\sigma_{peak} = \sigma_c - \frac{a_2}{d^2} = \sigma_c - A(\theta - \theta_0)^2
\label{eqn:ThetaDependenceOfStress}
\end{equation}
where $a_2$ combines the three contributions described above. This parabolic
dependence on $\theta-30^{\circ}$ is depicted on the left-hand side of
figure~\ref{fig:HighAnglePeakStress}.

% There is an
%abrupt jump down in peak stress as we move away from the perfect crystal because
%adding dislocations to the perfect crystal adds a nucleation site for fracture.

\subsection{High Angle Grain Boundaries}
\label{sec:HighAnglePeakStress}
\begin{figure}[thb]
\begin{center}
\includegraphics[width=\figWidth]{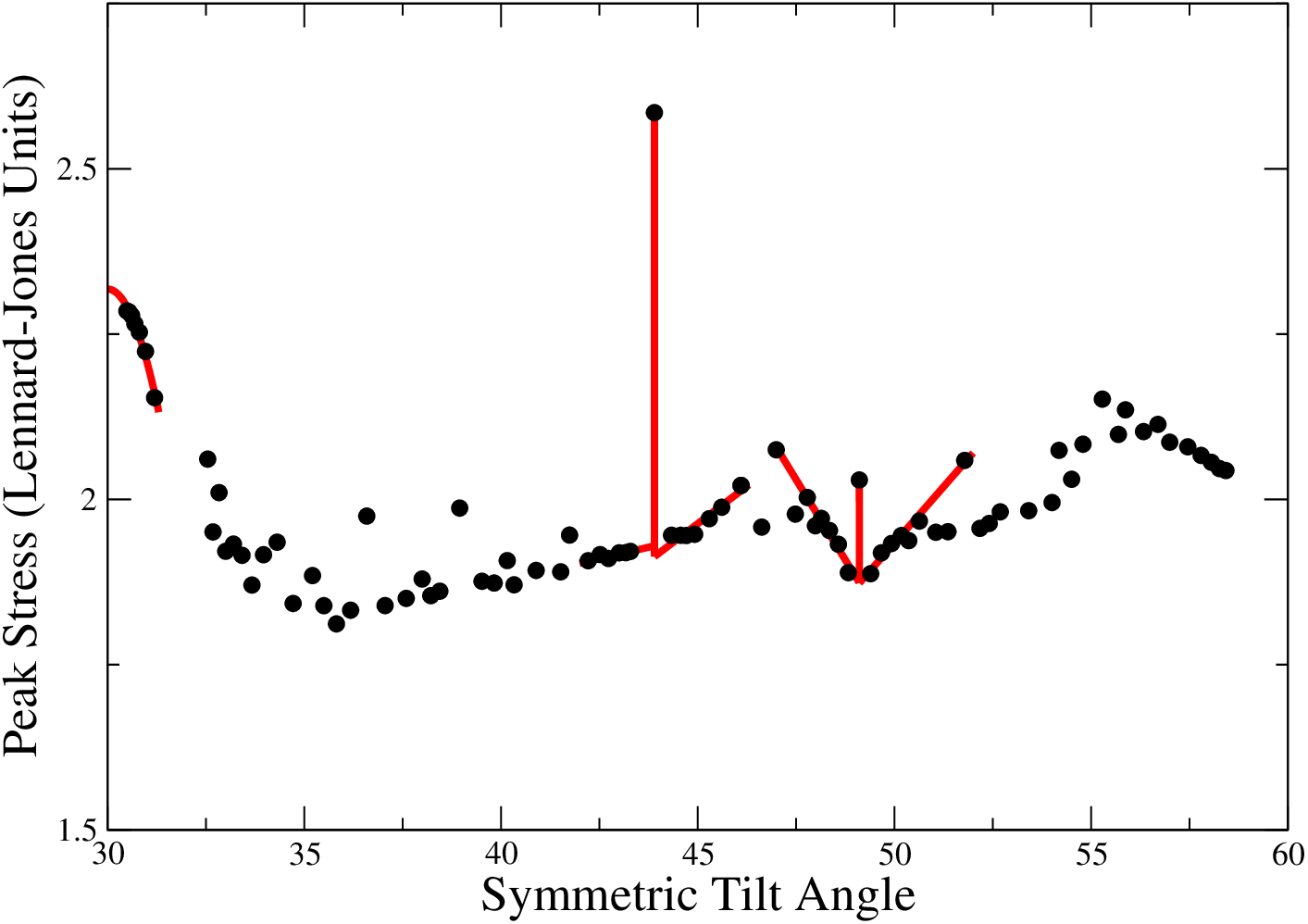}
\end{center}
\mycaption[Peak Stress vs. Tilt Angle For High Angle Grain
Boundaries.]{\coloronline The
peak stress as a function of tilt angle is discontinuous everywhere, with higher
values at special tilt angles representing high symmetry grain boundary
geometries.  The dependence of peak stress on angle near the high symmetry grain
boundaries (parabolic near 30$^{\circ}$, linear near 49.11$^{\circ}$) depends on
the structure of the additional flaws that make up the nearby geometries.}
\label{fig:HighAnglePeakStress}
\end{figure}

Figure~\ref{fig:HighAnglePeakStress} shows the results of the peak stress
measurements for high angle symmetric grain boundaries.  At the same points for
which we had cusps in energy, we have discontinuous increases in fracture
strength.  By drawing the same analogy between adding dislocations to perfect
crystals and adding flaws to high symmetry boundaries as described in figure
\ref{fig:ExtraDislocations} we can understand the discontinuities in the
fracture strength at high symmetry grain boundaries and the angular dependence
of fracture strength near the high symmetry geometries.

% the data for this plot is in the ccmr directory
% ASP/MDWebServices/GrainBreaker/SymmetricGrainBoundaries

\begin{figure}[th]
\begin{center}
\inThesis{
\includegraphics[height=5.5cm]{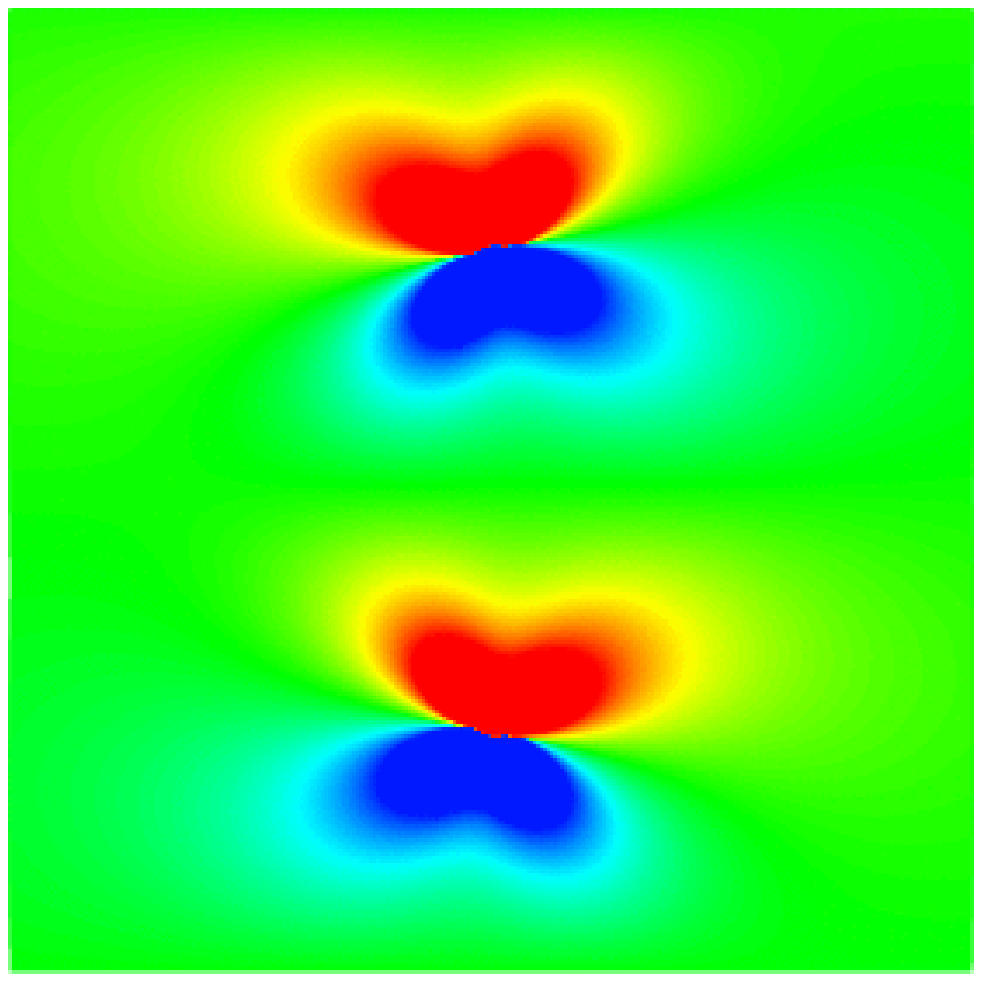}
\hskip .5cm
\includegraphics[height=5.5cm]{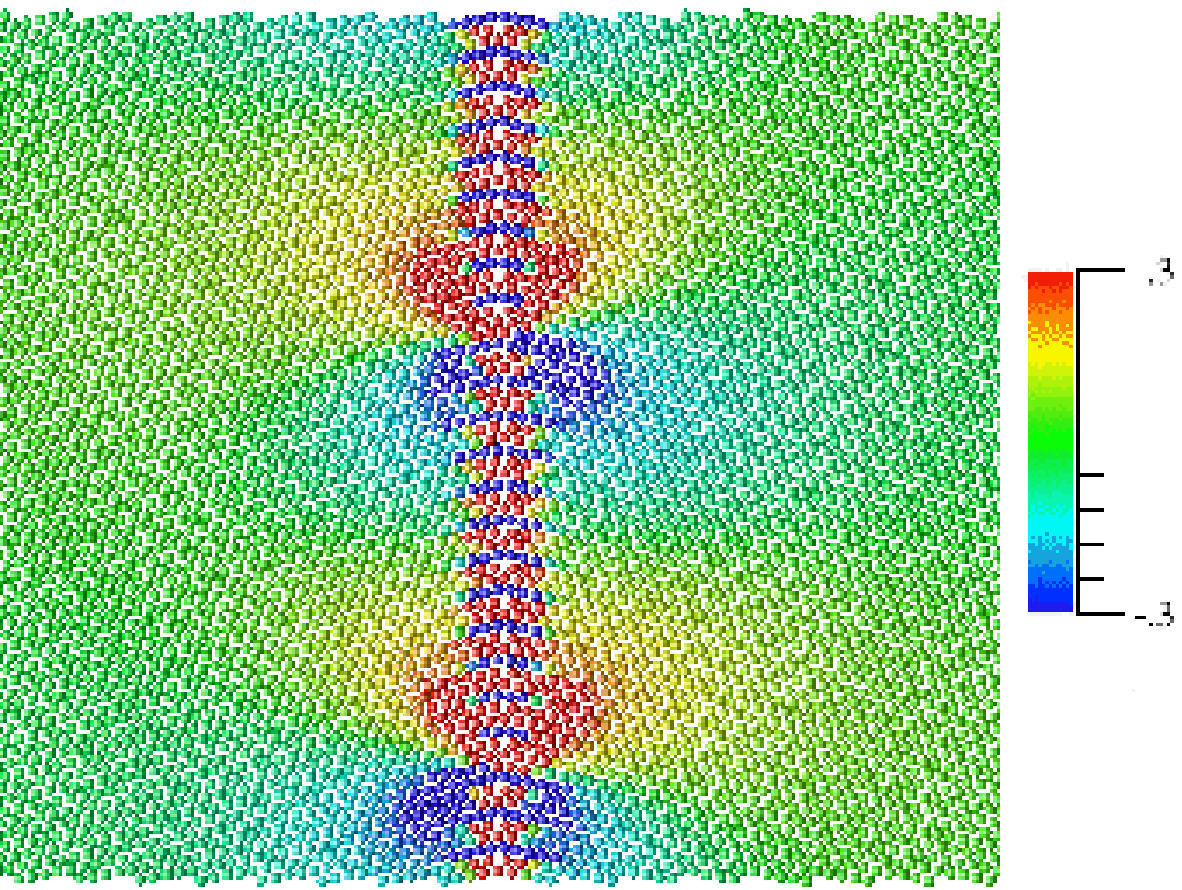}}
\inPaper{
\includegraphics[height=3.5cm]{grain_boundary_figures/ContinuumStressField}
\hskip .25cm
\includegraphics[height=3.5cm]{grain_boundary_figures/stressfield01color}}
\end{center}
\mycaption[ Stress Fields Due to Dislocations.]{\coloronline The figure on the left shows the
  stress fields surrounding two dislocations according to
  formula~\ref{eqn:VolterraSolution}.  
  The figure on the right shows the stress fields
  surrounding the added dislocations as calculated according to the virial
  definition of atomic stress~\cite{allen-tildesley}.}
\label{fig:StressFields}
\end{figure}

For high angle grain boundaries, the added flaw is no longer the sole nucleation
site for fracture and fracture does not necessarily nucleate in the core of the
added flaw.  The added dislocation creates a stress field given roughly by the
Volterra solution (equation~\ref{eqn:VolterraSolution}) with a positive stress
on one side, negative stress on the other, and a singularity at the center shown
in figure~\ref{fig:StressFields}.  The stress field differs slightly from the
Volterra solution because the elastic constants of the material at the grain
boundary vary from those of the perfect crystal.  The fracture nucleates along
the boundary in the region where the stress due to the added flaw is positive.
Because fracture does not nucleate at the center of the added flaw, the Volterra
solution as summed over the neighboring, added flaws does not cancel at the
nucleation site.  This leads to a linear law for fracture strength as a function
of tilt angle, for grain boundaries near high symmetry geometries.
%For high angle grain boundaries the crack does not nucleate at the extra flaws
%but in a region a few Angstroms above it along the grain boundary.  [show
%picture].  The position at which fracture is nucleated is uniform for similar
%flaw patterns (clarify this)In this case, the Volterra solutions due to the
%neighboring flaws do not cancel. \begin{figure}[thb]

% the data for this plot is in the ccmr directory
% ASP/MDWebServices/GrainBreaker/SymmetricGrainBoundaries/Base3+/geo123

% the data for this plot is in the ccmr directory
% ASP/MDWebServices/GrainBreaker/SymmetricGrainBoundaries
\begin{figure}[thb]
\begin{center}
\includegraphics[width=\figWidth]{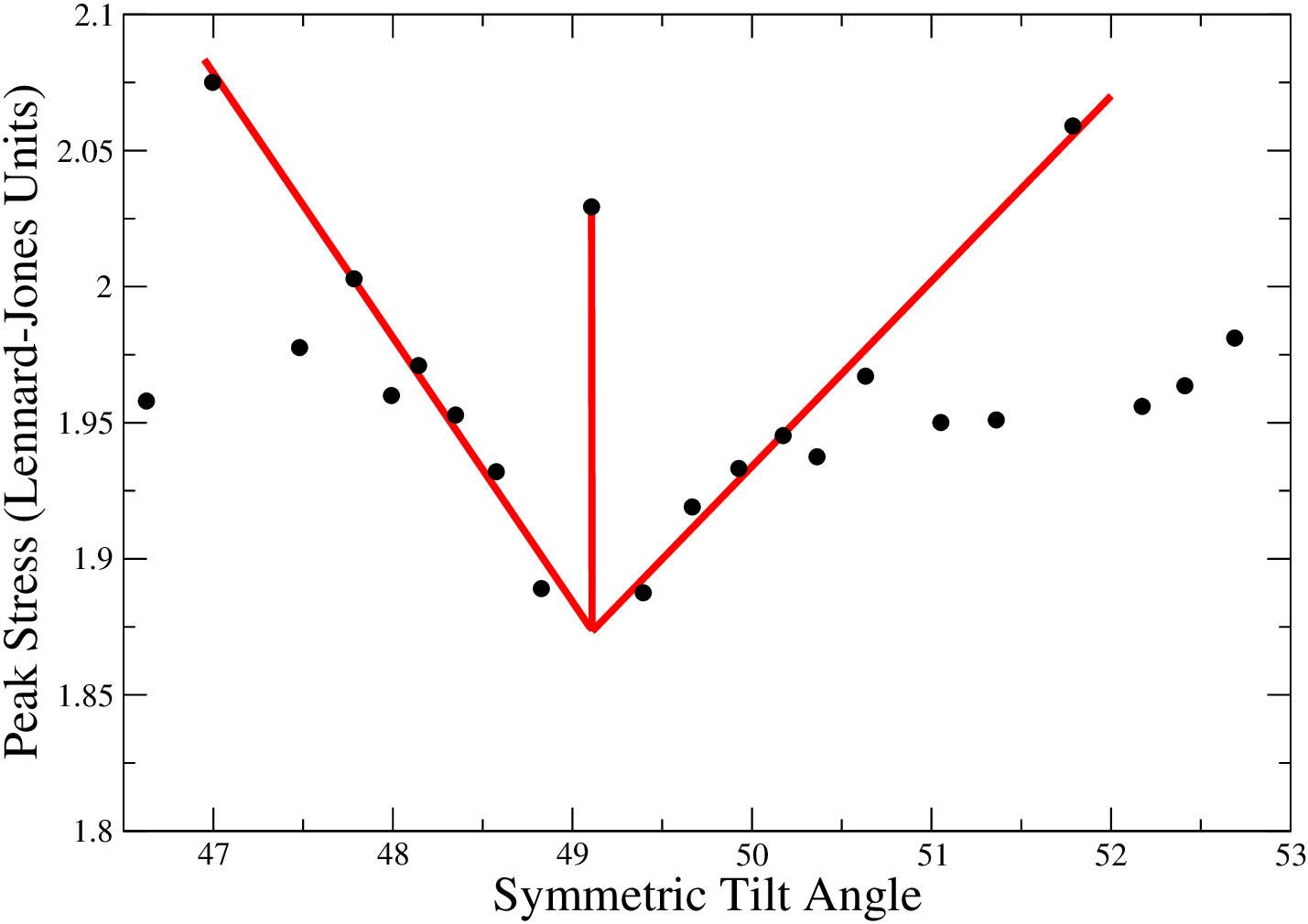}
\end{center}
\mycaption[Peak Stress vs. Tilt Angle Near High Symmetry Grain Boundaries.]{
\coloronline There
  is a discontinuity in peak stress at tilt angles close to high symmetry grain
  boundaries.  The plot above shows the peak stress for the grain boundary with
  tilt angle 49.1 (described in table~\ref{table:HighSymmetry}) and the nearby
  geometries.}
\label{fig:HighSymmetryPeakStress}
\end{figure}

Consider the grain boundaries with tilt angles ranging from 49.39$^{\circ}$ to
53.41$^{\circ}$, which are close to the high symmetry grain boundary at
49.11$^{\circ}$.  The additional flaws that characterize these grain boundaries
have Burger's vectors equal to $(-\sqrt{3} \sin \theta, -\sin \theta)$ and
$(-\sqrt{3} \sin \theta,\sin \theta)$, where $\theta$ is the symmetric tilt
angle of the lattice. The norms are $2 | \sin \theta|$ and the angles are 210
and 150$^{\circ}$.
%For a Burger's vector at angle $\alpha$ from the $x$-axis.
%\begin{equation}
%\sigma_{xx}(x,y) = \frac{\mu b}{2 \pi (1-\nu)} \frac{y(3 x^2 +
%    y^2) \cos \alpha - x (x^2 - y^2) \sin \alpha}{(x^2+y^2)^2}
%\end{equation}
For these geometries, the $xx$ component of the stress field (due to two
dislocations a distance $D \approx b/2(\theta_0-\theta)$) along the $y$-axis is
\begin{eqnarray}
\sigma_{xx}(y) &=& \frac{ \mu(2 \sqrt{3} y (\theta_0 - \theta) - 3 \sin
  (\theta_0 - \theta)) \sin (\theta_0 - \theta)} {2 \pi (1 - \nu) (y (\theta_0 -
  \theta) - \sqrt{3} \sin (\theta_0 - \theta))} \\ &\approx& \frac{\mu
  (3-2\sqrt{3} y) (\theta_0 - \theta)}{2 \pi y (1-\nu) (y-\sqrt{3})} +
O((\theta_0 - \theta))^3
\end{eqnarray}
where $\theta_0$ is the tilt angle of the high symmetry grain boundary.  We need
to look at simulations in the fixed displacement (narrow width) regime in order
to observe where fracture nucleates.~\footnote{For simulations of a few
angstroms wide, the crack opens up along the grain boundary with a slow
unzipping mechanism, starting at particular flaws.  For wider simulations, the
crack snaps open, soon after a peak stress is reached. The slow unzipping of the
narrow simulations is characteristic of fixed displacement boundary conditions -
since the displacements of a region close to the interface is fixed, the opening
of the crack is controlled, allowing us to see the details of how the crack
opens.  The snapping open, seen with a wider simulation, is characteristic of
fixed force boundary conditions. In the wider simulations, the material on
either side of the grain boundary acts as a spring, which effectively softens
the boundary conditions, approximating fixed-force boundaries for long lengths.
When the constrained region is far away, the interface snaps open once it
reaches the maximum stress that it can sustain.} We find that for geometries with this
pattern of flaws, fracture nucleates at the same distance above the added flaw.
The external stress needed to nucleate fracture at a distance y from the added
flaw, along the grain boundary is then approximately
\begin{equation}
\sigma_{peak} = \sigma_{c} -  \frac{\mu (3-2\sqrt{3} y) (\theta_0 - \theta)}{2 \pi y (1-\nu)
  (y-\sqrt{3})}
\end{equation}
which explains the linear dependence on angle shown in
figures~\ref{fig:StressFields} and~\ref{fig:HighSymmetryPeakStress}.

\begin{figure}[thb]
\begin{center}
\includegraphics[width=8.5cm]{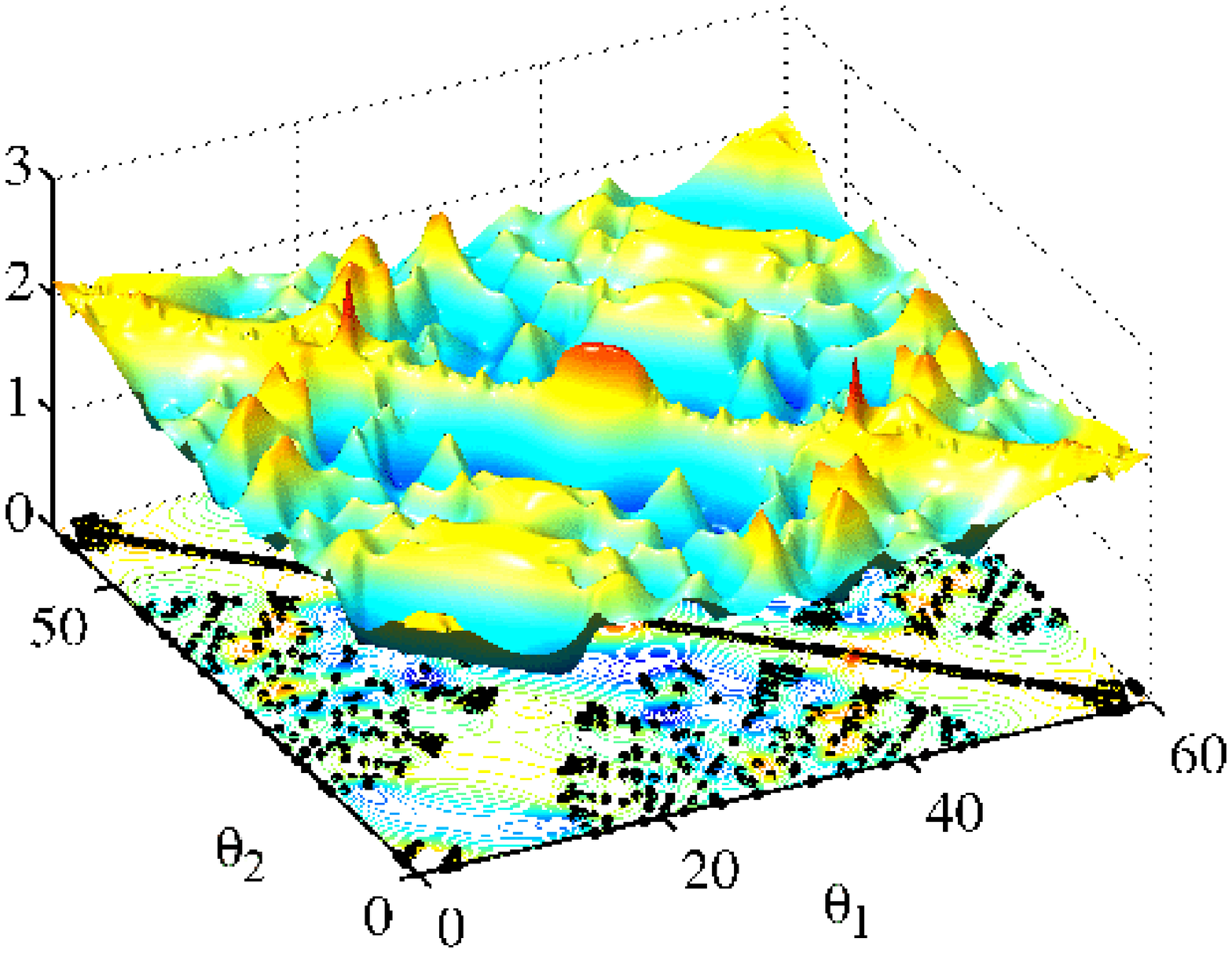}
\mycaption[Grain Boundary Peak Stresses for Asymmetric Geometries]{As with
figure~\ref{fig:AsymmetricEnergies}, the surface and contours represent a smooth
interpolation between particular geometries.   Due to the interpolation, each high
stress peak is rendered as a finite-width cone.  The peak stress of perfect crystals is
much higher than that of grain boundaries.  In order to use a smaller vertical
scale, we have not included these data points  which would
otherwise show a ridge along the diagonal.}
\label{fig:AsymmetricPeakStress}
\end{center}
\end{figure}

\begin{figure}[thb]
\begin{center}
\includegraphics[width=8.5cm]{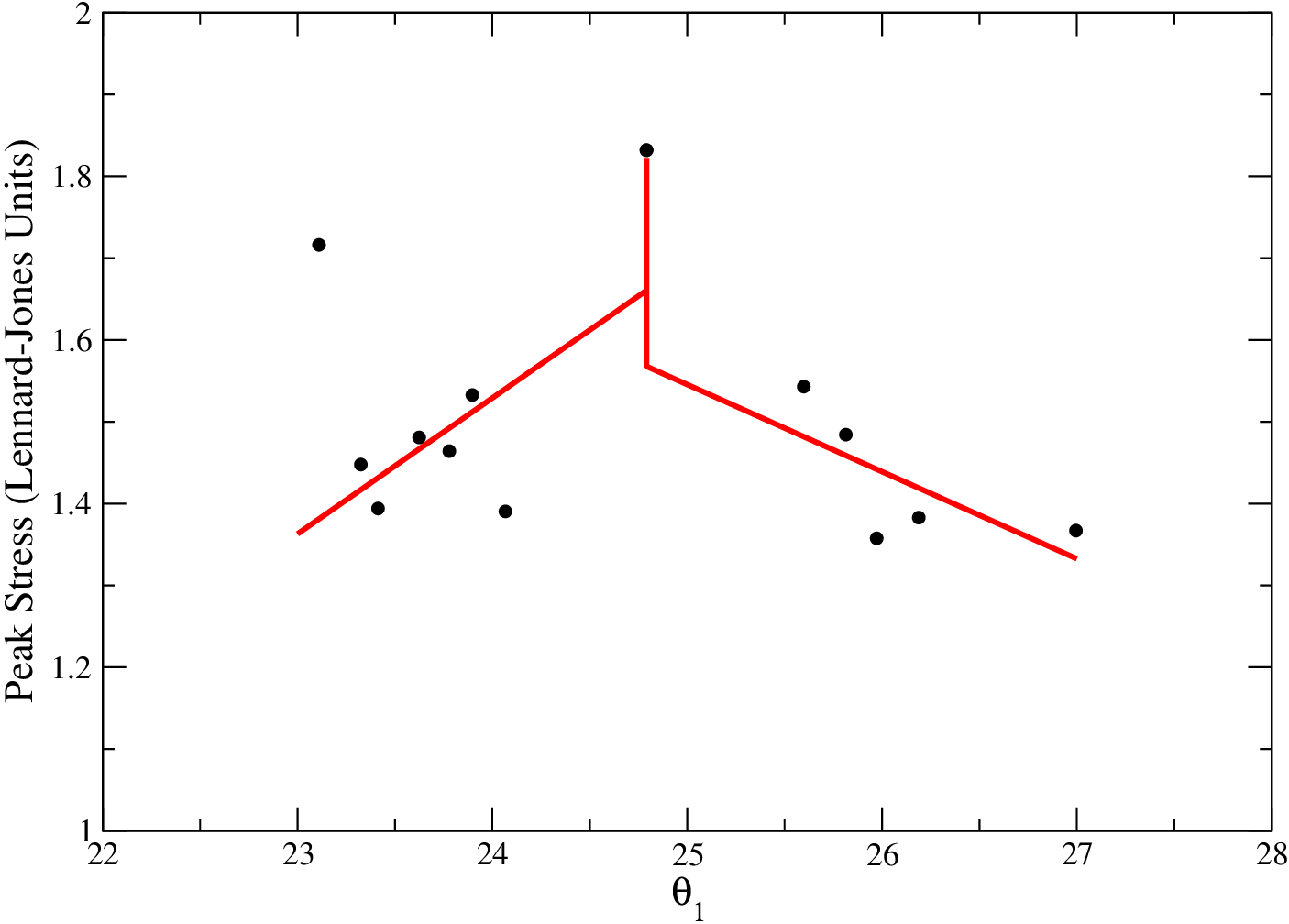}
\mycaption[Asymmectric Grain Boundary Peak Stresses In Neighborhood of High
Symmetry Geometry.]{The grain boundary geometries plotted above are the same as
those in figure~\ref{fig:AsymmetricEnergiesNeighborhood}, along a line of
constant misorientation in the vicinity of a high symmetry geometry.  The lines
show a possible linear fit to the points on either side of the low repeat
distance grain boundary.  }
\label{fig:AsymmetricPeakStressNeighborhood}
\end{center}
\end{figure}

The results for the peak stress of all grain boundaries with repeat lengths less
than 30 lattice constants are shown in figure~\ref{fig:AsymmetricPeakStress}.
We also note spikes in peak stress for grain boundaries with particularly low
repeat distances.  Consider again the grain boundary with tilt angles
24.79$^\circ$ and 3.00$^\circ$ which has a repeat distance of 9.53 lattice
constants.  Figure~\ref{fig:AsymmetricPeakStressNeighborhood} shows the peak
stress of this grain boundary and grain boundaries in its neighborhood, along a
line of constant misorientation.

\section{Conclusion}

We have systematically explored the space of 2D grain boundaries, the
patterns of structural units that make up short repeat distance grain
boundaries, and the flaws in the patterns of structural units that make up
vicinal grain boundaries. We have shown that the patterns of
structural units are key to understanding the singularities in energy and peak
stress at special grain boundaries by drawing an analogy to perfect crystals.
We have used this insight to find a functional form for the energies of 2D
symmetric grain boundaries and to understand the tilt angle dependence of peak
stress near special grain boundaries.

In principle, with enough computer time it would be possible to conduct a similar study for
the 5D space of idealized 3D grain boundaries.  We suspect that the perfect
crystal analogy could also explain the cusps in energy and discontinuous spikes
in peak stress for special 3D grain boundaries.  The geometry dependence of
energy and peak stress surrounding these special grain boundaries will be more
complicated since the grain boundaries can have a pure tilt, pure twist, or
a mixed type of geometry.  In real polycrystals, the fracture strength and energy
are further complicated by impuritites at the interface, emission of
dislocations during fracture, and more complex geometries such as triple
junctions of grains, which are often the nucleation site for fracture.
Systematic studies of such systems are infeasible and better suited by on the
fly simulations of local regions of interest.

\begin{acknowledgements}
This work was supported by NSF Grants No. ITR/ASP ACI0085969 and
No. DMR-0218475.  We also wish to thank Nicholas Bailey, Drew Dolgert, Gerd
Heber, Anthony Ingraffea, Surachute Limkumnerd, Chris Myers, and Paul Wawrzynek.
\end{acknowledgements}

\clearpage

\bibliographystyle{apsrev}
\bibliography{../general}

\begin{thebibliography}{25}
\expandafter\ifx\csname natexlab\endcsname\relax\def\natexlab#1{#1}\fi
\expandafter\ifx\csname bibnamefont\endcsname\relax
  \def\bibnamefont#1{#1}\fi
\expandafter\ifx\csname bibfnamefont\endcsname\relax
  \def\bibfnamefont#1{#1}\fi
\expandafter\ifx\csname citenamefont\endcsname\relax
  \def\citenamefont#1{#1}\fi
\expandafter\ifx\csname url\endcsname\relax
  \def\url#1{\texttt{#1}}\fi
\expandafter\ifx\csname urlprefix\endcsname\relax\def\urlprefix{URL }\fi
\providecommand{\bibinfo}[2]{#2}
\providecommand{\eprint}[2][]{\url{#2}}

\bibitem[{\citenamefont{Needleman}(1990)}]{needleman-interfaceCZM}
\bibinfo{author}{\bibfnamefont{A.}~\bibnamefont{Needleman}},
  \bibinfo{journal}{Journal of the Mechanics and Physics of Solids}
  \textbf{\bibinfo{volume}{38}}, \bibinfo{pages}{289} (\bibinfo{year}{1990}).

\bibitem[{\citenamefont{Falk et~al.}(2001)\citenamefont{Falk, Needle\-man, and
  Rice}}]{falk-critical}
\bibinfo{author}{\bibfnamefont{M.}~\bibnamefont{Falk}},
  \bibinfo{author}{\bibfnamefont{A.}~\bibnamefont{Needle\-man}},
  \bibnamefont{and} \bibinfo{author}{\bibfnamefont{J.}~\bibnamefont{Rice}}, in
  \emph{\bibinfo{booktitle}{Pro-ceed\-ings of the 5th Eu\-ro\-pean
  Me\-chan\-ics of Materials Con\-fer\-ence}} (\bibinfo{address}{Delft},
  \bibinfo{year}{2001}),
  \urlprefix\url{http://citeseer.ist.psu.edu/490781.html}.

\bibitem[{\citenamefont{Iesulauro et~al.}(2002)\citenamefont{Iesulauro,
  Ingraffea, Arwade, and Wawrzynek}}]{iesulauro1}
\bibinfo{author}{\bibfnamefont{E.}~\bibnamefont{Iesulauro}},
  \bibinfo{author}{\bibfnamefont{A.~R.} \bibnamefont{Ingraffea}},
  \bibinfo{author}{\bibfnamefont{S.}~\bibnamefont{Arwade}}, \bibnamefont{and}
  \bibinfo{author}{\bibfnamefont{P.~A.} \bibnamefont{Wawrzynek}}, in
  \emph{\bibinfo{booktitle}{Fatigue and Fracture Mechanics}}, edited by
  \bibinfo{editor}{\bibfnamefont{W.~G.} \bibnamefont{Reuter}} \bibnamefont{and}
  \bibinfo{editor}{\bibfnamefont{R.~S.} \bibnamefont{Piascik}}
  (\bibinfo{organization}{American Society for Testing and Materials},
  \bibinfo{address}{West Conshohocken, PA}, \bibinfo{year}{2002}),
  vol.~\bibinfo{volume}{33}.

\bibitem[{\citenamefont{Iesulauro et~al.}(2003)\citenamefont{Iesulauro,
  Ingraffea, Heber, and Wawrzynek}}]{iesulauro2}
\bibinfo{author}{\bibfnamefont{E.}~\bibnamefont{Iesulauro}},
  \bibinfo{author}{\bibfnamefont{A.~R.} \bibnamefont{Ingraffea}},
  \bibinfo{author}{\bibfnamefont{G.}~\bibnamefont{Heber}}, \bibnamefont{and}
  \bibinfo{author}{\bibfnamefont{P.~A.} \bibnamefont{Wawrzynek}}, in
  \emph{\bibinfo{booktitle}{44th AIAA/ASME/ASCE/AHS Structures, Structural
  Dynamics, and Materials Conference}} (\bibinfo{organization}{AIAA},
  \bibinfo{address}{Norfolk, VA}, \bibinfo{year}{2003}).

\bibitem[{\citenamefont{Warner et~al.}(2006)\citenamefont{Warner, Sansoz, and
  Molinari}}]{warner-sliding-plasticity}
\bibinfo{author}{\bibfnamefont{D.~H.} \bibnamefont{Warner}},
  \bibinfo{author}{\bibfnamefont{F.}~\bibnamefont{Sansoz}}, \bibnamefont{and}
  \bibinfo{author}{\bibfnamefont{J.~F.} \bibnamefont{Molinari}},
  \bibinfo{journal}{International Journal of Plasticity}
  \textbf{\bibinfo{volume}{22}}, \bibinfo{pages}{754} (\bibinfo{year}{2006}).

\bibitem[{\citenamefont{Coffman et~al.}(2007)\citenamefont{Coffman, Sethna,
  Heber, Liu, Ingraffea, and Barker}}]{cube-in-cube}
\bibinfo{author}{\bibfnamefont{V.~R.} \bibnamefont{Coffman}},
  \bibinfo{author}{\bibfnamefont{J.~P.} \bibnamefont{Sethna}},
  \bibinfo{author}{\bibfnamefont{G.}~\bibnamefont{Heber}},
  \bibinfo{author}{\bibfnamefont{A.}~\bibnamefont{Liu}},
  \bibinfo{author}{\bibfnamefont{A.}~\bibnamefont{Ingraffea}},
  \bibnamefont{and} \bibinfo{author}{\bibfnamefont{E.~I.} \bibnamefont{Barker}}
  (\bibinfo{year}{2007}), \bibinfo{note}{in preparation}.

\bibitem[{\citenamefont{Chen et~al.}(1989)\citenamefont{Chen, Srolovitz, and
  Voter}}]{chen-GB}
\bibinfo{author}{\bibfnamefont{S.~P.} \bibnamefont{Chen}},
  \bibinfo{author}{\bibfnamefont{D.~J.} \bibnamefont{Srolovitz}},
  \bibnamefont{and} \bibinfo{author}{\bibfnamefont{A.~F.} \bibnamefont{Voter}},
  \bibinfo{journal}{Journal of Materials Research}
  \textbf{\bibinfo{volume}{4}}, \bibinfo{pages}{62} (\bibinfo{year}{1989}).

\bibitem[{\citenamefont{Sansoz and
  Molinari}(2004)}]{sansoz-molinari-shear-tension}
\bibinfo{author}{\bibfnamefont{F.}~\bibnamefont{Sansoz}} \bibnamefont{and}
  \bibinfo{author}{\bibfnamefont{J.~F.} \bibnamefont{Molinari}},
  \bibinfo{journal}{Scripta Materialia} \textbf{\bibinfo{volume}{50}},
  \bibinfo{pages}{1283} (\bibinfo{year}{2004}).

\bibitem[{\citenamefont{Sansoz and Molinari}(2005)}]{sansoz-molinari-structure}
\bibinfo{author}{\bibfnamefont{F.}~\bibnamefont{Sansoz}} \bibnamefont{and}
  \bibinfo{author}{\bibfnamefont{J.~F.} \bibnamefont{Molinari}},
  \bibinfo{journal}{Acta Materialia} \textbf{\bibinfo{volume}{53}},
  \bibinfo{pages}{1931} (\bibinfo{year}{2005}).

\bibitem[{\citenamefont{Shenderova et~al.}(2000)\citenamefont{Shenderova,
  Brenner, Omeltchenko, Su, and Yang}}]{shenderova-diamond}
\bibinfo{author}{\bibfnamefont{O.~A.} \bibnamefont{Shenderova}},
  \bibinfo{author}{\bibfnamefont{D.~W.} \bibnamefont{Brenner}},
  \bibinfo{author}{\bibfnamefont{A.}~\bibnamefont{Omeltchenko}},
  \bibinfo{author}{\bibfnamefont{X.}~\bibnamefont{Su}}, \bibnamefont{and}
  \bibinfo{author}{\bibfnamefont{L.~H.} \bibnamefont{Yang}},
  \bibinfo{journal}{Phys. Rev. B} \textbf{\bibinfo{volume}{61}},
  \bibinfo{pages}{3877} (\bibinfo{year}{2000}).

\bibitem[{\citenamefont{Spearot et~al.}(2004)\citenamefont{Spearot, Jacob, and
  McDowell}}]{spearot-atomistic}
\bibinfo{author}{\bibfnamefont{D.~E.} \bibnamefont{Spearot}},
  \bibinfo{author}{\bibfnamefont{K.~I.} \bibnamefont{Jacob}}, \bibnamefont{and}
  \bibinfo{author}{\bibfnamefont{D.~L.} \bibnamefont{McDowell}},
  \bibinfo{journal}{Mechanics of Materials} \textbf{\bibinfo{volume}{36}},
  \bibinfo{pages}{825} (\bibinfo{year}{2004}).

\bibitem[{\citenamefont{Bollmann}(1970)}]{bollmann}
\bibinfo{author}{\bibfnamefont{W.}~\bibnamefont{Bollmann}},
  \emph{\bibinfo{title}{Crystal Defects and Crystalline Interfaces}}
  (\bibinfo{publisher}{Springer-Verlag}, \bibinfo{address}{New York},
  \bibinfo{year}{1970}).

\bibitem[{\citenamefont{Bishop and Chalmers}(1968)}]{bishop-chalmers}
\bibinfo{author}{\bibfnamefont{G.~H.} \bibnamefont{Bishop}} \bibnamefont{and}
  \bibinfo{author}{\bibfnamefont{B.}~\bibnamefont{Chalmers}},
  \bibinfo{journal}{Scripta Metallurgica} \textbf{\bibinfo{volume}{2}},
  \bibinfo{pages}{133} (\bibinfo{year}{1968}).

\bibitem[{\citenamefont{Ishida}(1976)}]{bubble-raft}
\bibinfo{author}{\bibfnamefont{Y.}~\bibnamefont{Ishida}},
  \emph{\bibinfo{title}{Grain Boundary Structure and Properties}}
  (\bibinfo{publisher}{Academic Press Inc.}, \bibinfo{address}{London},
  \bibinfo{year}{1976}), chap. \bibinfo{chapter}{The Bubble Raft as a Model for
  Grain Boundary Structure}, pp. \bibinfo{pages}{93--106}.

\bibitem[{\citenamefont{Wolf and Merkle}(1992)}]{wolf-merkle}
\bibinfo{author}{\bibfnamefont{D.}~\bibnamefont{Wolf}} \bibnamefont{and}
  \bibinfo{author}{\bibfnamefont{K.~L.} \bibnamefont{Merkle}},
  \emph{\bibinfo{title}{Materials Interfaces: Atomic-level structure and
  properties}} (\bibinfo{publisher}{Chapman \& Hall}, \bibinfo{year}{1992}),
  chap. \bibinfo{chapter}{Correlation between the structure and energy of grain
  boundaries in metals}, pp. \bibinfo{pages}{88--150}.

\bibitem[{\citenamefont{Harrison et~al.}(1976)\citenamefont{Harrison,
  Bruggeman, and Bishop}}]{old-grain-boundary-sims}
\bibinfo{author}{\bibfnamefont{R.~J.} \bibnamefont{Harrison}},
  \bibinfo{author}{\bibfnamefont{G.~A.} \bibnamefont{Bruggeman}},
  \bibnamefont{and} \bibinfo{author}{\bibfnamefont{G.~H.}
  \bibnamefont{Bishop}}, \emph{\bibinfo{title}{Grain Boundary Structure and
  Properties}} (\bibinfo{publisher}{Academic Press Inc.},
  \bibinfo{address}{London}, \bibinfo{year}{1976}), chap.
  \bibinfo{chapter}{Computer Simulation Methods applied to Grain Boundaries},
  pp. \bibinfo{pages}{45--91}.

\bibitem[{\citenamefont{Pumphrey}(1976)}]{pumphrey}
\bibinfo{author}{\bibfnamefont{P.~H.} \bibnamefont{Pumphrey}},
  \emph{\bibinfo{title}{Grain Boundary Structure and Properties}}
  (\bibinfo{publisher}{Academic Press Inc.}, \bibinfo{address}{London},
  \bibinfo{year}{1976}), chap. \bibinfo{chapter}{Special High Angle Grain
  Boundaries}, pp. \bibinfo{pages}{139--200}.

\bibitem[{\citenamefont{Wolf}(1992)}]{wolf-structure}
\bibinfo{author}{\bibfnamefont{D.}~\bibnamefont{Wolf}},
  \emph{\bibinfo{title}{Materials Interfaces: Atomic-level structure and
  properties}} (\bibinfo{publisher}{Chapman \& Hall}, \bibinfo{year}{1992}),
  chap. \bibinfo{chapter}{Atomic-level geometry of crystalline interfaces}, pp.
  \bibinfo{pages}{1--52}.

\bibitem[{\citenamefont{Wolf and Jaszczak}(1992)}]{wolf-jaszczak}
\bibinfo{author}{\bibfnamefont{D.}~\bibnamefont{Wolf}} \bibnamefont{and}
  \bibinfo{author}{\bibfnamefont{J.~A.} \bibnamefont{Jaszczak}},
  \emph{\bibinfo{title}{Materials Interfaces: Atomic-level structure and
  properties}} (\bibinfo{publisher}{Chapman \& Hall}, \bibinfo{year}{1992}),
  chap. \bibinfo{chapter}{Role of interface dislocations and surface steps in
  the work of adhesion}, pp. \bibinfo{pages}{662--690}.

\bibitem[{\citenamefont{Hirth and Lothe}(1982)}]{HirthAndLothe}
\bibinfo{author}{\bibfnamefont{J.~P.} \bibnamefont{Hirth}} \bibnamefont{and}
  \bibinfo{author}{\bibfnamefont{J.}~\bibnamefont{Lothe}},
  \emph{\bibinfo{title}{Theory of Dislocations}} (\bibinfo{publisher}{John
  Wiley \& Sons}, \bibinfo{address}{New York}, \bibinfo{year}{1982}).

\bibitem[{\citenamefont{Bailey et~al.}(2006)\citenamefont{Bailey, Cretegny,
  Sethna, Coffman, Dolgert, Myers, Schiotz, and Mortensen}}]{DigitalMaterial}
\bibinfo{author}{\bibfnamefont{N.}~\bibnamefont{Bailey}},
  \bibinfo{author}{\bibfnamefont{T.}~\bibnamefont{Cretegny}},
  \bibinfo{author}{\bibfnamefont{J.~P.} \bibnamefont{Sethna}},
  \bibinfo{author}{\bibfnamefont{V.~R.} \bibnamefont{Coffman}},
  \bibinfo{author}{\bibfnamefont{A.~J.} \bibnamefont{Dolgert}},
  \bibinfo{author}{\bibfnamefont{C.~R.} \bibnamefont{Myers}},
  \bibinfo{author}{\bibfnamefont{J.}~\bibnamefont{Schiotz}}, \bibnamefont{and}
  \bibinfo{author}{\bibfnamefont{J.~J.} \bibnamefont{Mortensen}},
  \emph{\bibinfo{title}{Digital material: a flexible atomistic simulation
  code}} (\bibinfo{year}{2006}),
  \urlprefix\url{http://arxiv.org/abs/cond-mat/0601236}.

\bibitem[{\citenamefont{Palumbo and Aust}(1992)}]{palumbo-aust}
\bibinfo{author}{\bibfnamefont{G.}~\bibnamefont{Palumbo}} \bibnamefont{and}
  \bibinfo{author}{\bibfnamefont{K.~T.} \bibnamefont{Aust}},
  \emph{\bibinfo{title}{Materials Interfaces: Atomic-level structure and
  properties}} (\bibinfo{publisher}{Chapman \& Hall}, \bibinfo{year}{1992}),
  chap. \bibinfo{chapter}{Special Properties of $\Sigma$ grain boundaries.},
  pp. \bibinfo{pages}{190--207}.

\bibitem[{\citenamefont{Bailey}(2002)}]{NicksThesis}
\bibinfo{author}{\bibfnamefont{N.}~\bibnamefont{Bailey}}, Ph.D. thesis,
  \bibinfo{school}{Cornell} (\bibinfo{year}{2002}).

\bibitem[{\citenamefont{Allen and Tildesley}(1987)}]{allen-tildesley}
\bibinfo{author}{\bibfnamefont{M.~P.} \bibnamefont{Allen}} \bibnamefont{and}
  \bibinfo{author}{\bibfnamefont{D.~J.} \bibnamefont{Tildesley}},
  \emph{\bibinfo{title}{Computer Simulation of Liquids}}
  (\bibinfo{publisher}{Oxford Science Publications}, \bibinfo{year}{1987}).

\bibitem[{\citenamefont{Frenkel and T.}(1938)}]{frenkel-kontorova}
\bibinfo{author}{\bibfnamefont{Y.~I.} \bibnamefont{Frenkel}} \bibnamefont{and}
  \bibinfo{author}{\bibfnamefont{K.}~\bibnamefont{T.}},
  \bibinfo{journal}{Zhurnal tekhnicheskoi fiziki} \textbf{\bibinfo{volume}{8}},
  \bibinfo{pages}{1340} (\bibinfo{year}{1938}).

\end{thebibliography}

\end{document}